\documentclass{llncs}
\usepackage{graphicx}
\usepackage{amsmath}
\usepackage{caption}
\usepackage{capt-of}

\usepackage[caption = false]{subfig}

\usepackage{algorithm}
\usepackage{verbatim}
\usepackage[noend]{algpseudocode}
\usepackage{amssymb}

\usepackage{lipsum}
\usepackage{picins}

\DeclareCaptionFormat{algor}{%
	\hrulefill\par\offinterlineskip\vskip1pt%
	\textbf{#1#2}#3\offinterlineskip\hrulefill}
\DeclareCaptionStyle{algori}{singlelinecheck=off,format=algor,labelsep=space}
\begin{document}
\title{Scalable Top-$k$  Query on Information Networks with Hierarchical Inheritance Relations}
%

\author{\vspace*{-1mm} 
	Fubao Wu, Lixin Gao}
%
%
\institute{University of Massachusetts, Amherst  \\
\email{fubaowu@umass.edu}   \\
}
\maketitle              
\vspace*{-4mm}
\begin{abstract}
Graph query, pattern mining and knowledge discovery become challenging on large-scale heterogeneous information networks (HINs). State-of-the-art techniques involving path propagation mainly focus on the inference on nodes labels and neighborhood structures. However, entity links in the real world also contain rich hierarchical inheritance relations. For example, the vulnerability of a product version is likely to be inherited from its older versions. Taking advantage of the hierarchical inheritances can potentially improve the quality of query results. Motivated by this, we explore hierarchical inheritance relations between entities and formulate the problem of graph query on HINs with hierarchical inheritance relations. We propose a graph query search algorithm by decomposing the original query graph into multiple star queries and apply a star query algorithm to each star query. Further candidates from each star query result are then constructed for final top-$k$ query answers to the original query. To efficiently obtain the graph query result from a large-scale HIN, we design a bound-based pruning technique by using uniform cost search to prune search spaces.
We implement our algorithm in GraphX to test the effectiveness and efficiency on synthetic and real-world datasets. Compared with two common graph query algorithms, our algorithm can effectively obtain more accurate results and competitive performances.

\end{abstract}
\vspace*{-7mm}
\section{Introduction}
\vspace*{-2.3mm}
%

Many real-world systems, such as enterprise networks, social networks, and biological networks, can be modeled as heterogeneous information networks (HIN)  \cite{shi2017survey}. HIN contains multiple types of objects and relations providing rich semantic queries, knowledge discoveries, information fusions, recommendations and predictions. Graph query, as an important technique for solving these tasks, has been extensively explored recently. It mainly explores subgraph isomorphism algorithms to get an exact match \cite{fujiwara2012fast}, and also develop subgraph matching algorithms to do an inexact/approximate match as the potential query answers \cite{khan2013nema,jin2015querying}. Current research on graph query/matching mainly focuses on two dimensions. The first dimension is the unary node-to-node properties mapping. The second dimension is edge-to-edge/path similarities. Jin et al. \cite{jin2015querying} consider node types and closest path propagations to get scores of query answers. Some works \cite{khan2013nema,yang2014schemaless} consider similar nodes' labels and their neighbors to learn the path propagation to get ranked answers.

However, knowledge representation has hierarchical structures in the real world system. Long et al. state that the knowledge structure representation can be inherited with upward and downward inheritances \cite{long1988inheritance}. Clauset et al. show that the existing knowledge of hierarchical structure can be used to predict missing connections \cite{clauset2008hierarchical}. In addition, Jiang et al. construct the hierarchical structures of entities for the large freebase knowledge base system based on real world entities and relations \cite{jiang2015entity}. One visible example in an enterprise's product databases is that product vulnerabilities can be inherited from or passed down to different product versions. While measuring the similarity of objects for graph matching, hierarchical inheritance relations can also play an important role for the answer ranking. The quality of query answers is also affected by hierarchical inheritance relations. Therefore, we consider the power of hierarchical inheritances whereby a subclass inherits properties and constraints of its parents, and more meaningful and accurate query answers are expected to be obtained.

\begin{figure}[!ht]
	\subfloat[]{\includegraphics[width= 1.3in, height=1.1in]{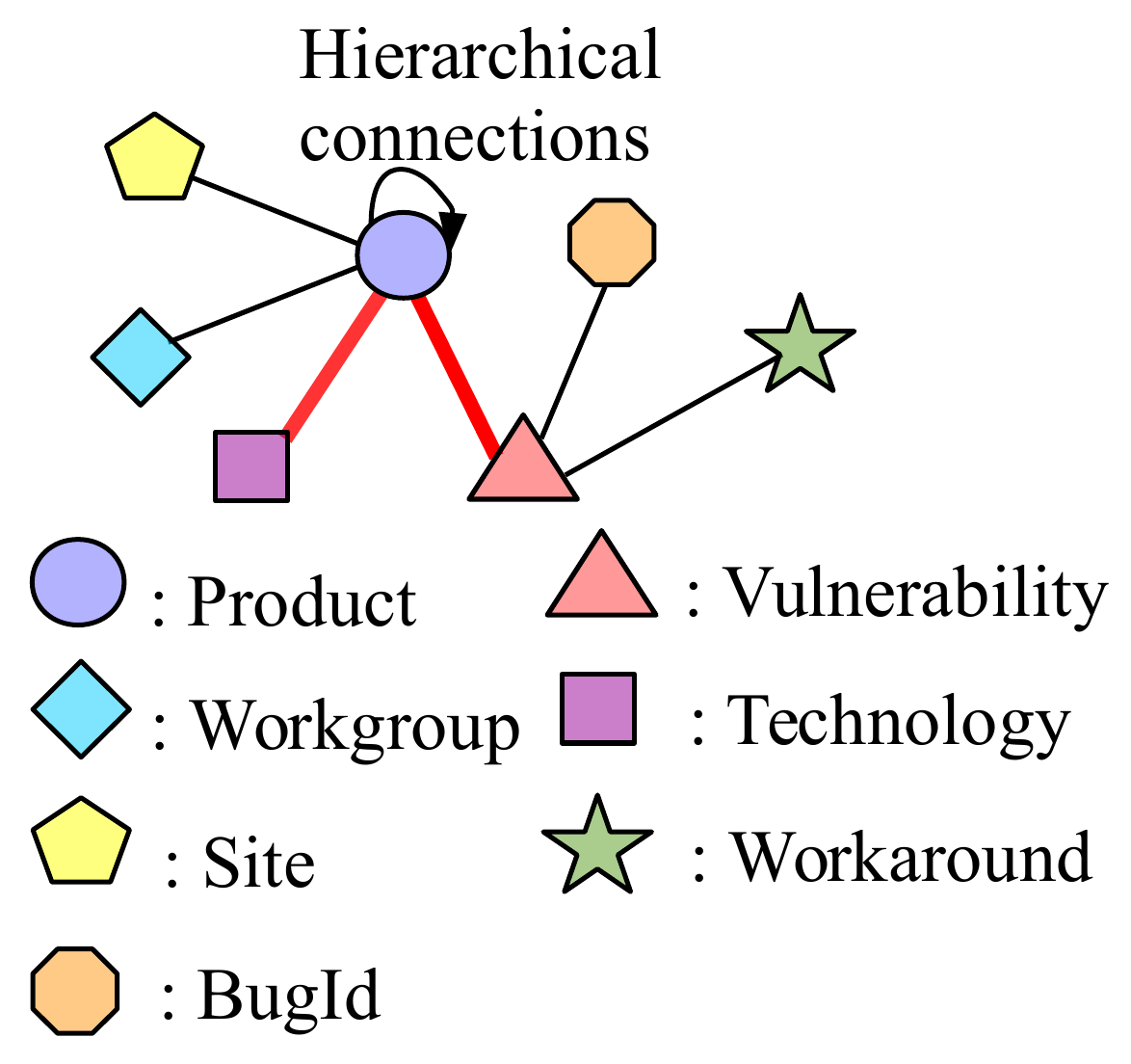}} \label{fig:introGraphSchemaExample01}
	\subfloat[]{\includegraphics[width= 1.4in, ,height=1.1in]{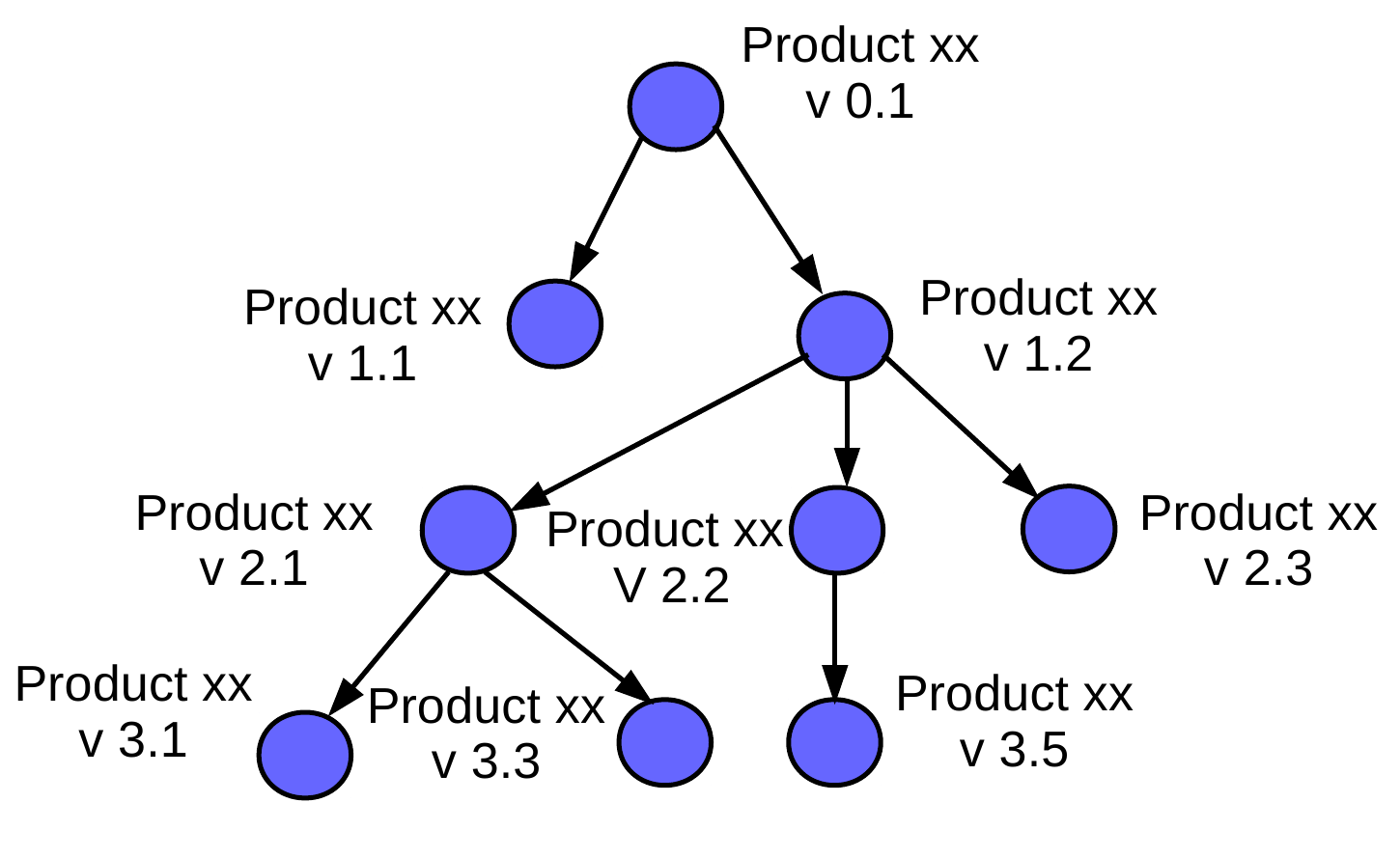}} \label{fig:introGraphSchemaHierarchical01} 	\hspace{2mm}
	\subfloat[]{\includegraphics[width= 0.9in, height=0.7in]{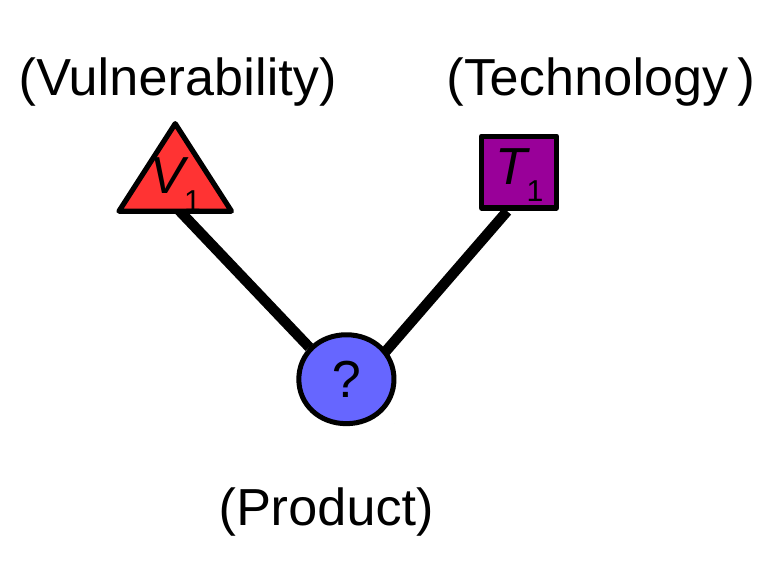}} \label{fig:introExampleStarQuery01}
	\subfloat[]{\includegraphics[width= 1.0in, ,height=1.1in]{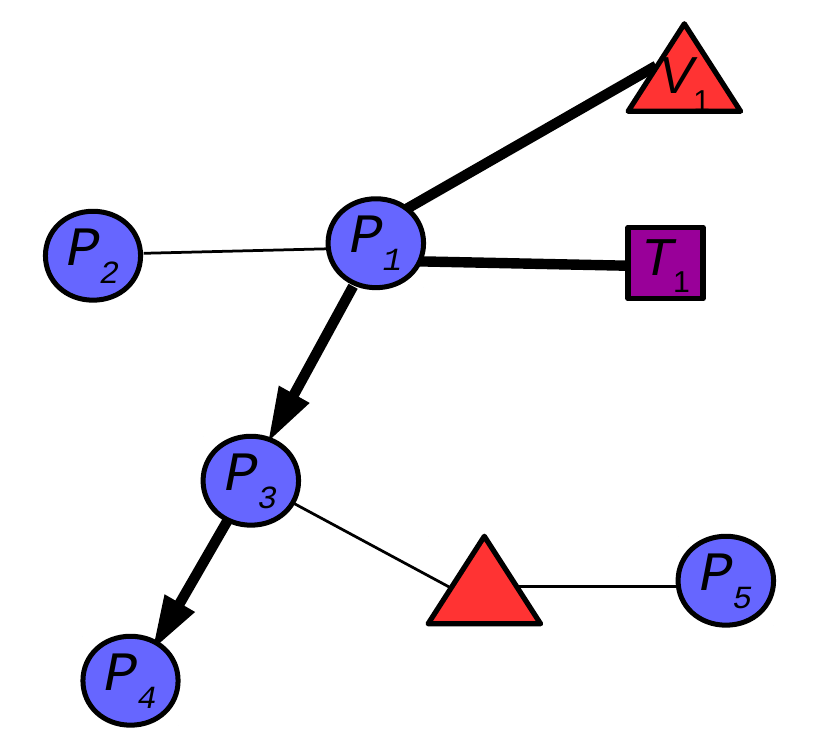}} \label{fig:introExampleStarQueryResult01} 
	\vspace*{-2mm}
	\caption{A product information network schema and a query example. (a) A graph schema of a product network  (b) An example of a hierarchical connection structure for product versions (c) A user query graph of top products (d) A answer graph with top-$5$ products}  
	\vspace*{-4mm}
	\label{fig:querySchemaExamples}
\end{figure}

Taking an example of an information network with hierarchical structure, we consider a schema of an enterprise's product information network shown in Fig \ref{fig:querySchemaExamples}a. Every node represents a type of entity at the schema level. The product type is connected by four property types: site, workgroup, technology and vulnerability. Product entities have hierarchical connections with different versions of products shown in Fig \ref{fig:querySchemaExamples}b. Some properties are inherited among different versions of products, such as vulnerability and technology properties (in red bold lines in Fig \ref{fig:querySchemaExamples}a).

Given the information network schema with inherited relations, we show a user query example here. Assume a user wants to find the top-$5$ related products affected with a given vulnerability $V_1$ (Cisco WebEx meetings server information disclosure vulnerability) and employed with a given technology $T_1$ (voice - communications manager additional apps and plugins), which is constructed as a user query graph shown in Fig \ref{fig:querySchemaExamples}c. Fig \ref{fig:querySchemaExamples}d shows a possible top-$5$ subgraph answers of the query in this answer graph. The given $V_1$ and $T_1$ node in user query graph is exactly matched with the $V_1$ and $T_1$ node respectively in the answer graph, and there are 5 product nodes which are possible answers to the query of product node.

For general methods, if we consider closest node types and shortest distances to measure the similarities of answers for matching, we get the following ranking order of answer scores, $P_1$ (Cisco WebEx meetings server versions 0.1.0), $P_2$ (Cisco WebEx meetings server versions 0.2.0), $P_3$ (Cisco WebEx meetings server versions 1.1), $P_4$ (Cisco WebEx meetings server versions 2.1) and $P_5$ (Cisco Jabber for Windows), that is, the ranking order of answer scores is $s(P_1) > s(P_2) = s(P_3) > s(P_4) > s(P_5)$. However, the vulnerability property can be inherited from different prior versions of products. Here $P_1$ is the prior (parent) version of $P_3$, and $P_3$ is the parent version of $P4$ as the arrow indicates. Hence, $P_1$'s vulnerabilities can pass down to product $P_3$ or $P_4$, and $P_4$'s vulnerability can come from upper $P_3$ or $P_1$. With the hierarchical inheritances, the answer scores can have a more accurate ranking order $s(P_1) \approx s(P_3) \approx s(P_4) > s(P_2) > s(P_5)$, which is important for engineers' troubleshooting and customers' queries.

\vspace*{-0.5mm}
Due to the complexity and heterogeneousness of large networks, designing an effective and efficient algorithm with additional hierarchical features is challenging. In this paper, we conquer this problem by modeling graph query with a new matching score function with hierarchical inheritance relations for effective answers, and by proposing a bound-based technique for an efficient query. The main contributions are as follows:
\vspace*{-2mm}
\begin{itemize}
	\item We formulate the graph query problem with hierarchical inheritance relations to improve query quality.
	\item We propose a new graph query algorithm based on uniform cost search in the context of a new matching score function.
	\item We design a bound-based method to prune search spaces to efficiently get the top-$k$ best answers.
	\item We implement our algorithm in the Spark GraphX distributed environment for large-scale networks. Experiments are done to evaluate the effectiveness and efficiency of our matching algorithm.
\end{itemize}
The rest of this paper is organized as follows. Section \ref{problemForm} describes the problem and formulates graph queries with hierarchical inheritance relations. The proposed algorithm for graph queries and its bound-based pruning technique are presented in Section \ref{algorithm}. Section \ref{implementation} discusses the distributed implementation. In Section \ref{evaluation}, we present the evaluation of our algorithms. The related work and conclusion are shown in Sections \ref{relatedWork} and \ref{conclusion}.

\vspace*{-1mm}
\section{Problem Formulation}
\vspace*{-2mm}
\label{problemForm}
\subsection{Data Graph, Query Graph and Matching}
\vspace*{-2.3mm}
We consider a HIN that contains hierarchical inheritance relations among nodes as a hierarchical heterogeneous information network (HHIN). HHIN is modeled as a partially undirected, labeled data graph $G(V,E,L_v,H_e)$ with a node set V, edge set E, node label set $L_v$ and hierarchical inheritance relations $H_e$ with directions, where (1) each node $v \in V$  represents an entity in $G$, (2) each edge $e \in E$ represents the relationship between two entities, and each edge weight is considered to be 1. Only an edge between two hierarchical entities has a direction. (3) each node $v$ has a label information $L_v$, including at least a node type and a keyword description, (4) especially, for hierarchical entities, each edge $e \in H_e$ between them indicates a hierarchical inheritance relation. Each edge weight between two hierarchical entities is $|H_e| = 1$ .

There exists upward and downward hierarchical inheritance relations in $G$. We call a node with label information that is inherited among other node's hierarchical entities as an ``attaching'' node, such as a vulnerability node whose label information that could be inherited among product nodes in Fig \ref{fig:querySchemaExamples}a. A node with a node type that has hierarchical levels is called an ``inherited'' node, such as a product node in Fig \ref{fig:querySchemaExamples}a. If an attaching node's label information passes down to its inherited node's lower level entity, we call it downward inheritance. Conversely, if an attaching node's label information can pass up to its inherited entity's higher level entity, it is called upward inheritance. The attaching node and one of its inherited nodes are formed as a ``property inheritance pair''. For example, the vulnerability's label information in the vulnerability entity can be downward or upward inherited from product entities in higher or lower levels as shown in red bold line in Fig \ref{fig:querySchemaExamples}a. The label information of workgroup or site is not inherited among product nodes as shown in black line in Fig \ref{fig:querySchemaExamples}a.

A query graph $Q(V_Q,E_Q,L_v)$ is modeled as an undirected and labeled graph. $V_Q$ contains a set of specific nodes $V_Q^S$ and a set of query nodes $V_{Q}^{U}$ with types $\tau_{Q}^{U}$, which are provided by users. A specific node is defined as an instantiated node in $Q$ that has a fixed node type and node label information, and it is also matched to a node in $G$. A query node is defined as a node in $Q$ given with node type only that we want to find its matched nodes in $G$. According to one category based on query node number, if the query node number $|V_{Q}^{U}| = 1$,  we denote the query graph as a star query graph. If the query node number $|V_{Q}^{U}| > 1$, it is called a general (non-star) query graph. According to another category based on hierarchical inheritance relations, if each of the specific nodes in $V_Q^S$ can form a property inheritance pair with any one of query node in $V_{Q}^{U}$, we call it a hierarchical query graph. If there exists at least one property inheritance pair between them, we call it a mixed hierarchical query graph. Otherwise, it is called a non-hierarchical query graph. For example, Fig \ref{fig:querySchemaExamples}c shows a hierarchical star query graph where node $V_1$ and node $T_1$ comprise specific nodes, and the node marked with ``?'' with product type represents a query node.

Given a query graph $Q$ and a data graph $G$, we need to map each query node to a data node. This transfers to a subgraph matching problem. We denote as $M$ an already matched subgraph in $G$ to $Q$. Then a subgraph matching is a many/one-to-one mapping function $\phi$: $V_Q \rightarrow V$, such that, for each query node $v \in V_Q$, $\phi(v) \in M$. The problem here is to find such top-$k$ potential mapping functions given a query graph $Q$ and a data graph $G$.
\vspace*{-2.8mm}
\subsection{Matching Score} 
\vspace*{-1.8mm}
If nodes are close in a query graph, their mapping nodes in a data graph are also close based on node neighbors and hierarchical inheritance relations.
Given a query graph $Q$ containing a node pair $(u, v) \in V_Q$ that is connected, a matched subgraph $M$ in $G$ has mapped nodes $(\phi(u), \phi(v))$.
\vspace*{-2.8mm}
\subsubsection{Node closeness score:} We define node closeness score based on whether hierarchical inheritances exist in $Q$.

(1) When $u$ and $v$ do not form a property inheritance pair, the closeness score of  $(\phi(u), \phi(v))$ is defined as:
\begin{equation}
\label{nodeSim}
r(\phi(u),\phi(u)) = \begin{cases}
1 & \text{ if } \phi(u)=\phi(v) \\ 
\alpha^{l(\phi(u),\phi(v))} & \text{ otherwise}
\end{cases}
\end{equation}
where  $l(\phi(u), \phi(v)$ is the shortest distance from $\phi(u)$ to $\phi(v)$. $\alpha$ is a constant propagation factor that controls the decreasing rate of node closeness within the value in $[0, 1]$. 

(2) When $u$ and $v$ can form a property inheritance pair, the closeness score of $(\phi(u), \phi(v))$ is defined as:
\begin{equation}
\label{nodeSim}
r(\phi(u),\phi(v)) = \begin{cases}
1 & \text{ if } \phi(u) = \phi(v) \\ 
\alpha^{l(\phi(u),\phi(v)-\beta*|h(\phi(u), \phi(v)|} & \text{ otherwise}
\end{cases}
\end{equation}

where  $l(\phi(u), \phi(v)$ is the shortest distance from $\phi(u)$ to $\phi(v)$. $\alpha$ is a constant propagation factor in $[0, 1]$ that controls the decreasing rate of node closeness. $\beta$ is defined as the hierarchical level propagation factor in $(0, 1)$, which indicates the importance of hierarchical level propagation when an attaching node's label information inherits between different hierarchical levels of an inherited node. $\beta$ is expected to be smaller than $\alpha$ because hierarchical inheritance is more reliable than shortest distances when traversing long hops. $h(\phi(u),\phi(v))$ indicates the hierarchical level difference from $\phi(u)$ to $\phi(v)$. Vice versa, the hierarchical level difference from $\phi(v)$ to $\phi(u)$ is indicated as $h(\phi(v), \phi(u))$,  and $ h(\phi(v), \phi(u)) = -h(\phi(u), \phi(v))$.

Based on the node closeness score, the matching score of $M$ is defined as the summation of mapping nodes $(\phi(u), \phi(v))$ for all connected edges $(u, v)$ in $Q$.

\begin{equation}
\label{equ:matchingScoreGeneralQuery}
F(\phi) = {\sum_{(u, v) \in E_Q}r(\phi(u), \phi(v))}
\end{equation}

\vspace*{-4.8mm}
\subsection{Problem statement:}
Given a query graph $Q$ and a data graph $G$, we want to find the top-$k$ subgraph answers in $G$, that is, to find a set of $k$ subgraphs $M_k$ in $G$, such that for any nodes $\phi (V_Q)\in M_k$ and for all nodes $\phi'(V_Q)\notin M_k$, the matching score $F(\phi) > F(\phi')$. Specific nodes $V_{Q}^{S}$ in $Q$ are identified to exactly one-to-one mapping to matched nodes $\phi(V_{Q}^{S})$ in $G$ (we call them anchor nodes $V_{G}^{A}$), which are easily to be found. Therefore, we consider the top-$k$ sets of candidate nodes in $M_k$ for a set of query nodes based on hierarchical inheritance relations and graph structures.

Formally, given a query graph $Q(V_Q,E_Q,L_v)$, the top-$k$ subgraphs $M_k(V', E', L')$ have the following mapping function with $Q$.
For each $v \in V_Q$, there is a one-to-one mapping $\phi(v) \in V'$:
$v\rightarrow \phi(v)$ based on the matching score $F$. Because our problem considers exact and approximate matches to output the top-$k$ matching answers, the edge of $e \in E_Q$ does not need to have a one-to-one mapping to the edge $e' \in E'$.

\vspace*{-2.8mm}
\section{Graph Query Algorithm with Hierarchical Inheritance Relations}
\vspace*{-1.8mm}
\label{algorithm}
It is time-consuming to get all potential subgraphs from a large-scale data graph with a big query graph.
Also, for a general query graph with multiple query nodes, it is proved to be an NP-hard problem even for subgraph isomorphism \cite{lee2012depth}. Yang et al. \cite{yang2016fast} divide a query graph into star query and then utilize the top-$k$ star-join method, using the similar relational database HRJN \cite{ilyas2004supporting}. Inspired by the structure of our general query graph with multiple query nodes, we propose a  general graph query algorithm comprising three phases as follows:

\textbf{Phase 1 (Query Decomposition)}: 
A general query graph contains some specific nodes and one or more query nodes. Considering the characteristics of our query graph, the decomposing policy of a general query graph is not as complex as the decomposing method considered in \cite{yang2016fast}, as we don't use the join for final combinations of star queries. Therefore, a simple and effective policy is to use the number of query nodes as the number of star query graphs. Each query node is the center query node for each star query, every specific node that is connected to the center query node is a specific node for its star query.
	
\textbf{Phase 2 (Star query)}:
We propose to use uniform cost search and bound-based pruning to derive top $k_s$ candidates for each star query. Selecting the top $k_s$ candidates for each star query can effectively serve the final top-$k$ candidate results for a general graph query (Section \ref{MatchingScoreStarQuery} -- \ref{starQueryAlgSection}).

\textbf{Phase 3 (Candidates selection)}:
We consider the top $k_s$ star query candidates together and get the optimum edge/path matching scores for query node combinations. Different from top-$k$ join strategy with a common node, without a common node for joining, query node candidates can be 1 or more hops connected in $G$. Therefore, graph traversals are needed among these star query node candidates to find the final top-$k$ candidate sets for query nodes. When there are $|V_{Q}^{U}|$ query nodes, this involves exponential $|{V_Q^U}|^{k_s}$ computations, which is highly expensive if $|{V_Q^U}|$ and $k_s$ are large. We propose to use a branch and bound technique to greatly reduce search spaces by filtering out unexpected candidate sets (Section \ref{candidateSelectionBounds}).

If the input query graph $Q$ is a star query graph, then we only do phase 2 (star query) to get the answer.
If the input query graph $Q$ is a general query graph, it will involve the three phases. As query decomposition is easy to accomplish, we will mainly discuss star query algorithm and candidate selection for general query graph algorithm.

\vspace*{-2.8mm}
\subsection{Matching Score for Star Query}
\label{MatchingScoreStarQuery}
Given a star query graph $Q$ with a set of specific nodes $V_{Q}^{S}$ and a query node $v$, specific nodes $V_{Q}^{S}$ have mapped to anchor nodes $V_{G}^{A}$ in $G$, so we only need to find the top-$k$ mapping nodes $\phi(v)$ for $v$. We denote $S(\phi(v))$ as the matching score of $\phi(v)$ based on the aggregated results of node closeness scores from all nodes in $\phi(V_{Q}^{S})$:

\begin{equation}
\label{matchingScoreEquation01}
S(\phi(v)) = \sum_{v^s \in V_{Q}^{S}}r(\phi(v^s), \phi(v)) 
\end{equation}

\vspace*{-5.8mm}
\subsection{Bound-based Pruning for Star Query}
\label{boundsStarQuery}
For each different anchor node, there is a propagation path to each candidate node in $G$. It is time-consuming to do all the node traversals if $G$ is very large. We use bound-based pruning technique to effectively reduce search spaces for star queries. We trace the lower bound in the top-$k$ answers and infer the upper bound of unseen nodes to effectively filter these nodes while traversing.
\vspace*{-3.8mm}
\subsubsection{Bounds of Matching Score}
In the top-$k$ answer list of query nodes, every node is maintained with a upper bound of matching score and a lower bound of matching score for a query node.
We refine the upper bound $\overline{S}(\phi(v))$ and lower bounds $\underline{S}(\phi(v))$. 
\begin{equation}
\label{equ:boundMatchingScore01}
\overline{S}(u) = \sum_{v^s \in V_{Q}^{S}}\overline{r}(\phi(v^s), \phi(v))
\end{equation}
\begin{equation} 
\label{equ:boundMatchingScore02}
\quad \underline{S}(u) = \sum_{v^s \in V_{Q}^{S}}\underline{r}(\phi(v^s), \phi(v))
\end{equation}

The matching score bound is dependent on the upper bound of node closeness score $\overline{r}$ and lower bound of closeness score  $\underline{r}$. 
Next, we show how to get these bounds of node closeness score.
\vspace*{-2.8mm}
\subsubsection{Bounds of Node Closeness Score}
The lower bound and upper bound are obtained online while the graph traversal is operated. We show the lower and upper bound refinement in the different iterations of graph traversal.
We denote $t$ as the iteration number of uniform cost search from an anchor node $s$ to a candidate node $u$.  

(1) The initial bounds is set as $\overline{r}^0(s,u) = 1 \quad and \quad \underline{r}^0(s,u) = 0$.
(2) In each of next iterations, every node $u$ is updated with its lower bound using the information from its previous iteration result when it is not visited yet. The lower bound is computed as follows:

\begin{equation}
\label{equ:boundIteration01}
\underline{r}^t(s,u) = \\
\left\{\begin{matrix}
\underline{r}^{t-1}(s,u) & \qquad \underline{r}^{t-1}(s,u) > 0  \\ 
\hspace{0.2in}  \alpha^{1-\beta\cdot|h(u_{prev}, u)|}\cdot\underline{r}^{t-1}(s, u_{prev})   & \qquad otherwise
\end{matrix}\right. 
\end{equation}
The upper bound in iteration $t$ is computed as follows:

\begin{equation}
\label{equ:boundIteration02}
\overline{r}^t(s,u) = \left\{\begin{matrix}
\underline{r}^{t}(s,u) & \qquad \underline{r}^{t}(s,u) > 0 \\ 
\alpha ^{t-\beta\cdot |h(s,u)|} & \qquad otherwise
\end{matrix}\right.
\end{equation}
where $u_{prev}$ is the parent node of $u$ when traversing from $s$ along a path to $u$.

\vspace*{-3.8mm}
\subsection{Top-$k$ Selection with Bounds}
\vspace*{-1.7mm}
\label{topKSelectionPruning}
How to effectively update potential candidate results and select the final top-$k$ results during the iterations is crucial for computation performances. Here we use a top-$k$ selection policy based on the upper and lower bounds of matching scores referred as the top-k emergence test in \cite{khemmarat2016fast}. We maintain a top-$k$ candidate result in a priority queue $P$. Each candidate node $u$ contains its lower bound $\overline{S}_G(u)$, and upper bound $\underline{S}_G(u)$. We define $\underline{S}_{kth}$ as the smallest lower bound in $P$. The process for selecting and updating $P$ during the iterations is shown as follows:

\textbf{(1)} Find the top-$k$ potential answer nodes and put in $P$.
\textbf{(2)} Calculate the ${kth}$ smallest lower bounds $\underline{S}_{kth}$ in $P$.
\textbf{(3)} If the upper bound $\overline{S}_G(u)$ of an incoming node $n$ is less than the $\underline{S}_{kth}$, we prune the node $u$ and the nodes with bigger distance than $u$ from the starting source. Because these nodes' matching scores are lower than any node's matching score in $P$, they are not qualified for top-$k$ final results. \textbf{(5)} Continue the previous steps until the convergence condition is reached shown in Section \ref{convergenceIteration}.

\vspace*{-3.8mm}
\subsection{Convergence of Iteration Propagation}
\vspace*{-1.7mm}
\label{convergenceIteration}
Two types of iteration conditions are identified to terminate the graph propagation to obtain the final top-$k$ answers. 

(1) When all the nodes with designated query node types have been explored or pruned by the bound-based pruning technique (Section \ref{topKSelectionPruning}), all the visited candidate nodes have obtained the necessary matching scores. 

(2) When no message updated for the next propagation, that is, all the candidate nodes' matching scores keep the same as the last iteration.

\vspace*{-3.8mm}
\subsection{Star Query Algorithm}
\vspace*{-1.3mm}
\label{starQueryAlgSection}
According to the proposed star query matching score and bounding-based pruning, we show our \underline{s}tar \underline{q}uery with \underline{h}ierarchical inheritance relations algorithm (SQH) in algorithm \ref{alg:starQueryAlgCode}. First, we get anchor nodes $\phi(V_Q^S)$ in $G$ for each $V_Q^S$ in $Q$, which are specific one-to-one mappings in $G$ (in Line \ref{alg:getnodeset}). Then we aggregate node messages to do propagation simultaneously from each anchor nodes with uniform cost search (in Line \ref{alg:aggregatenode}). Search cost of each node in uniform cost search is indicated by the inverse of its matching score here. In each iteration of propagation, the candidate node closeness and matching score, lower bounds and upper bounds are updated (in Lines \ref{alg:updatebound1}--\ref{alg:updatebound2}). Candidate nodes and the queue are continuously updated (in Lines \ref{alg:updateLM1}--\ref{alg:updateLM2}). The specific top-$k$ selection and update are shown (in lines \ref{alg:topkupdate1}--\ref{alg:topkupdate2}). Iterations continue until we found the final top-$k$ candidate result. The worst time complexity is $O(|V|*|V_Q^S|)$, where $|V|$ is the node number of $G$. With the pruning of potential unmatched nodes, the average time complexity is reduced to $O(M*|V_Q^S|)$, where $M$ is the number of visited nodes with type $\tau$ and $M \ll |V|$.
\vspace*{-2.8mm}

\lipsum*[0]
\begin{center}
	\label{alg:SQH}
	\captionof{algorithm}{Top-$k$ star query (SQH)}\label{alg:starQueryAlgCode}
	\begin{algorithmic}[1]
		\Require: Data graph $G(V,E,L_v, H_e)$, Query Graph $Q(V_Q^S,\tau)$, Matching number $k$
		\Ensure Top-$k$ match set $P_k$   
		\State  Get anchor nodes set $\phi(V_Q^S)$ for $V_Q^S$  \label{alg:getnodeset}
		\State Initialize empty match set $P_k$ (size $k$)  
		\State Initialize node closeness score  $r(s,u),\overline{r}(s,u)$ and $\underline{r}(s,u)$
		\State Initialize matching score $(S_G(u), \overline{S}_G(u), \underline{S}_G(u))$  
		
		\State Initialize $L \gets  \left \{  v|type(v) = \tau \hspace{1mm} \& \hspace{1mm} v \in V \right \}$
		\State $t \gets 0$
		\While{$L$ not empty and message exists}
		\State Aggregate node $u$ from each anchor nodes with uniform cost search  \label{alg:aggregatenode}  
		\State Update $(r(s, u), \overline{r}(s, u),   \underline{r}(s,u))$ by equation \ref{equ:boundIteration01} and \ref{equ:boundIteration02}   \label{alg:updatebound1}
		\State Update $(S_G(u), \overline{S}_G(u),   \underline{S}_G(u))$ by equations \ref{equ:boundMatchingScore01}  and \ref{equ:boundMatchingScore02} \label{alg:updatebound2}
		
		
		\If {$ \overline{S}_G(u) - \underline{S}_G(u)) <= 0 $} \label{alg:updateLM1}
		\State $L \gets L - u$   
		\Else
		\State $P_k = P_k + u$
		\EndIf 
		\State  $P_k, L \gets \textbf{TOPKUPDATEBOUNDPRUNE}$   \label{alg:updateLM2}
		\State $t \gets t + 1$
		\EndWhile
		\Procedure{$\textbf{TOPKUPDATEBOUNDPRUNE}$}{}  \label{alg:topkupdate1}
		\State  $\underline{S}_{kth}$ $\gets k_{th}$ smallest $\underline{S}_G(u)$  for $ u \in P_k$
		\State node n, ${S}_{kth}$ $\gets k_{th}$ smallest ${S}_G(u)$  for $ u \in P_k$
		\ForAll{$u \in P_k$}        
		\If {$size(P_k) < k$}
		\State $P_k \gets P_k + u $
		
		\ElsIf{$\overline{S}_G(u) > \underline{S}_{kth}(n)$ 
			\hspace{0.0in} and $S_G(u) > {S}_{kth}$}
		\State $P_k \gets P_k - n $
		\State $P_k \gets P_k + u $
		\ElsIf {$\overline{S}_G(u) < \underline{S}_{kth}(n)$}
		\State {$L \gets L - u$}    
		
		\EndIf
		\EndFor 
		\State \Return $P_k, L$      
		\EndProcedure      \label{alg:topkupdate2}
	\end{algorithmic}
\end{center}

\vspace*{-4.8mm}
\subsection{General Graph Query Algorithm}
The general graph query problem involves three phases described in the earlier part of Section \ref{algorithm}: decomposing query (phase 1),  star query (phase 2) and candidates selection (phase 3). The previous 2 phases has been described before. For phase 3, how to effectively and efficiently select the top matching candidate sets from star query results involves effective candidate selections. We propose to find the top matching scores of query node combinations by propagations. First, we define the matching score of query nodes for general graph query.
\vspace*{-3.8mm}
\subsubsection{Matching Score of Query Nodes}
Based on the definition of matching score for star query in Section \ref{MatchingScoreStarQuery}, we define the matching score for a set of query nodes $V_Q^U$ as:

\begin{equation}
\label{equ:matchingScoreGeneralQuery}
F_G(V_Q^U) =  \sum_{v \in \phi(V_Q^U) }S_G(v)+ \sum_{(v_i, v_j) \in {E(V_Q^U)} } E_G(\phi(v_i), \phi(v_j))
\end{equation}

The summation comprises of two parts. The first part is the summation of matching scores of decomposed star queries. The second part is the summation of matching scores of edges/paths among the candidates of query nodes. 
\vspace*{-3.8mm}
\subsubsection{Algorithm Flow} 

We show our \underline{g}eneral \underline{q}uery with \underline{h}ierarchical inheritance relations (GQH) in algorithm \ref{alg:generalQueryAlgCode}. Phase 1 for decomposing query is shown in line \ref{alg:decomposeQueryStar}. Phase 2 for star query is shown in line \ref{alg:starquery1}--\ref{alg:starquery2}. The candidate selection (in lines \ref{alg:candiSel1}--\ref{alg:candiSel2}) continues propagating by uniform cost search from top candidates nodes and pruning with branch and bound until the top-$k$ candidate node set is found. The worst time complexity is $O(|V|*|V_Q^S| + |V|*|V_Q^U|^{k_s})$, where $|V_Q^S|$ is the maximum number of specific nodes for each query node in a query graph, and $k_s$ is the number of top-$k_s$ candidate results from each star query result. In our experiment, $k_s \in [k, 2k]$ is a good trade-off for efficiency and effectiveness. $|V|$ is the number of nodes in $G$.
With the pruning of potential unmatched nodes for phase 2 and phase 3, the worst time complexity is reduced to $O(M*|V_Q^S|  + N*|V_Q^U|^{k_s})$, where $M$ and $N$ are the numbers of visited nodes with type $\tau$ for phase 2 and phase 3, respectively.
\vspace*{-3.6mm}
\lipsum*[0]
\begin{center}
	\captionof{algorithm}{Top-$k$ general query (GQH)}\label{alg:generalQueryAlgCode}
	\begin{algorithmic}[1]
	\Require: Data graph $G(V,E,L_v, H_e)$, Query Graph $Q(V_Q^S,\tau)$, Matching final number $k$
	\Ensure Top-$k$ matched candidate sets $Mt$         
	\State Initialize  top-$k$ matched candidate sets $Mt$ $\gets \phi  $ (size $k$) 
	\State Initialize star query result list $stResList$  $\gets \phi  $  
	\State  Star query graph set $StarGraphSet$ $\gets$  Query Graph $Q$    \label {alg:decomposeQueryStar}
	\State  $k_s \gets [k, 2k]$
	\ForAll{$starGraph \in StarGraphSet$}      \label{alg:starquery1}  
	\State Top $k_s$ candidate result $starCand  \gets$ SQH (Algorithm 1)
	\State $stResList \gets stResList + starCand$      \label{alg:starquery2} 
	\EndFor
	\State $i \gets 0$        \label{alg:candiSel1}
	\While{$i < len(stResList)-1$}
	\State Traverse from $stResList[i] \rightarrow stResList[i+1]$
	\State Pruning nodes and path with bounds until top-$k$ candidate node sets found 
	\State $i  \gets i + 1$    \label{alg:candiSel2}
	\EndWhile
	\end{algorithmic}
	\vspace*{-4.1mm}
\end{center}
\vspace*{-1.1mm}
\subsubsection{Candidate Selections with Branch and Bound Pruning}
\label{candidateSelectionBounds}
The output for a star query graph is top-$k_s$ candidate nodes for each query node. The problem is how to efficiently connect the candidate nodes of star query results and pick the top-$k$ answers. If all the candidate nodes are explored for each candidate combination, the time complexity would be exponential. We consider the branch and bound pruning technique \cite{clausen1999branch} while traversing among these candidate nodes. To ensure the best quality of candidate selections, we sort each top-$k_s$ result of star query in Phase 2 in a non-descending order in separate lists. Then we search through each list from the top to do uniform cost search and construct a search tree. Each path along the root to the leaf node is a matched candidate set for query nodes. While searching from root to leaf, we check aggregated matching scores, lower and upper bounds along the path. Assume there are top-$k$ candidate node sets with the smallest lower bound score $\underline{F}_{k_{th}}$, by searching the next candidate node and getting its upper bound lower than $\underline{F}_{k_{th}}$, the node candidate and all the nodes of its subtree can be pruned. 

\vspace*{-1.7mm}
\section{Distributed Implementation}
\vspace*{-2.1mm}
\label{implementation}
To support large information networks, we implement our graph query algorithm in the framework GraphX, which is a distributed graph analytics platform built on Apache Spark \cite{xin2013graphx}. We define a global data structure Global Vertex State Table ($GT$) for each vertex stored in the Spark RDD data structure. $GT$ is a user-defined class type which can store the following hash mapping for each anchor nodes $v^a$: node type $\tau$, shortest distance $sd$, hierarchical level difference $hd$, node closeness score $r$, closeness score lower bound $\underline{r}$, closeness score upper bound  $\overline{r}$, etc. $GT$ values are updated in each iteration of propagation to efficiently decide the bounds for effective pruning of many useless node propagations.

\vspace*{-1.8mm}
\section{Experimental Evaluation}
\vspace*{-2.3mm}
\label{evaluation}
The experiments are designed to answer the questions as follows:
(1) \textbf{Effectiveness:}
How is the quality of our query algorithm for hierarchical query graph or mixed query graph? how is the query with hierarchical inheritance relations compared with state-of-the-art methods?
(2) \textbf{Efficiency:} How is the efficiency and scalability of our algorithm on one machine and multiple machines?
\vspace*{-4.8mm}

\vspace*{-0.8mm}
\subsection{Datasets}  
\vspace*{-1.3mm}

We use synthetic data graph, Cisco product data graph and extended DBLP data graph.
Table \ref{tbl:dataset} shows the data statistics for our experiments.
\begin{table}[th]
	\centering
	\caption{Data set statistics}
	\label{tbl:dataset}
	\resizebox{\columnwidth}{!}{
		\begin{tabular}[b]{p{4.9cm}p{1.5cm}p{1.5cm}p{1.2cm}p{4.8cm}} 
			\hline
			\textbf{ Dataset}   & $|\textbf{V}|$ & $|\textbf{E}|$ &
			\textbf{Avg. degree}  & \textbf{No. of Vertex Types 
				(Attaching + Inherited + Other)} \\ \hline
			Synthetic Graph Data (Synthetic) & 10M     &  6.54M   & 10 &  2+2+3     \\ \hline
			
			Cisco Product Data (Cisco) & 111347    &  666992   & 12 &  2+1+4     \\ \hline
			Extended DBLP Data (DBLP)   & 1.28M    &35.1M     & 58 &  1+2+9       \\ \hline
		\end{tabular}
	}
\end{table}
\textbf{(1)} Synthetic data graph: we randomly generate data graph and create 7 types of nodes. There are 2 attaching node types, 2 inherited node types and 3 other node types. \textbf{(2)} Cisco data graph: we extract the data from its official and related support websites about devices and device properties, etc. The constructed graph schema is shown in Fig \ref{fig:querySchemaExamples}a. Vulnerability and Technology are the attaching node types, Product is the inherited node type. \textbf{(3)} Extended DBLP data graph: it is the DBLP database \cite{ley2005dblp} extending the topics extracted from lists of computer science conferences and journal websites. Topic is the attaching node type that is inherited among the conference/journal, paper and people node types.

\vspace*{-2.8mm}
\subsection{Quality of Graph Query}
\begin{figure}[!ht]
	\subfloat[]{\includegraphics[width= 2.6in, height=1.2in]{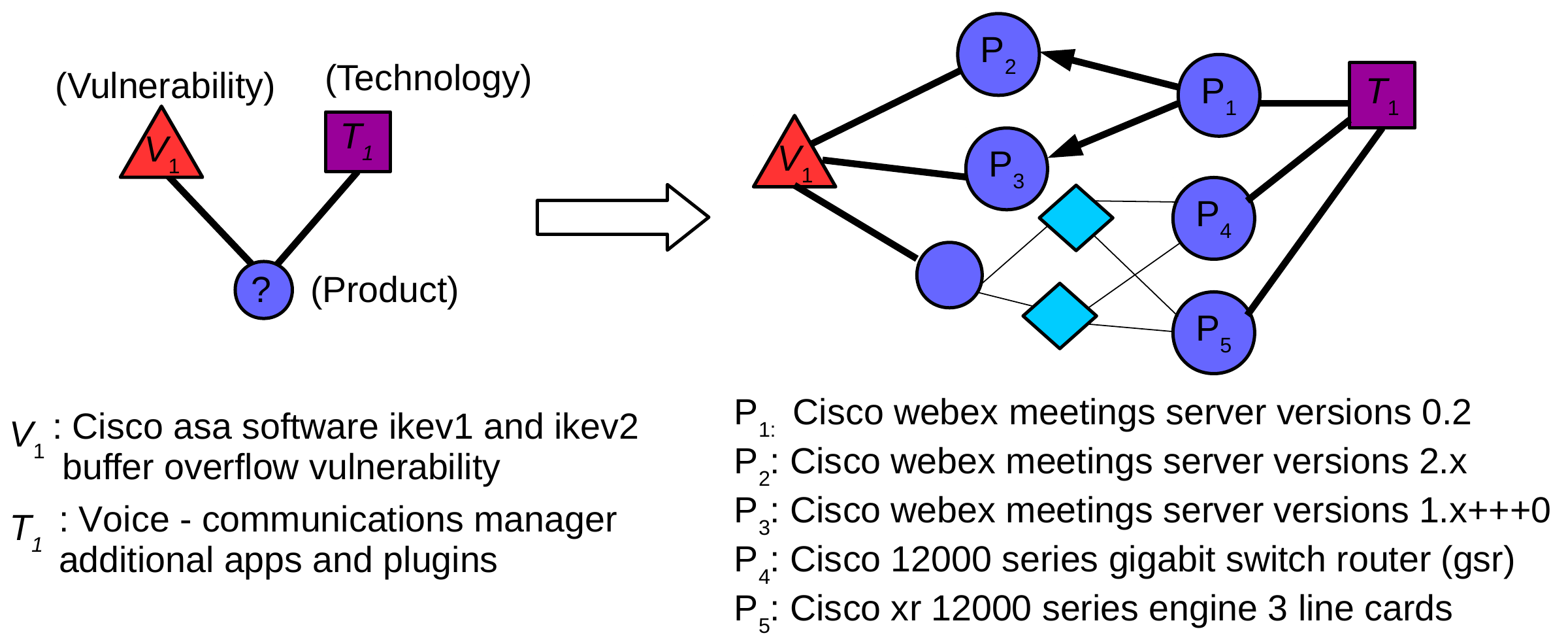}}
	\subfloat[]{\includegraphics[width= 2.6in, ,height=1.2in]{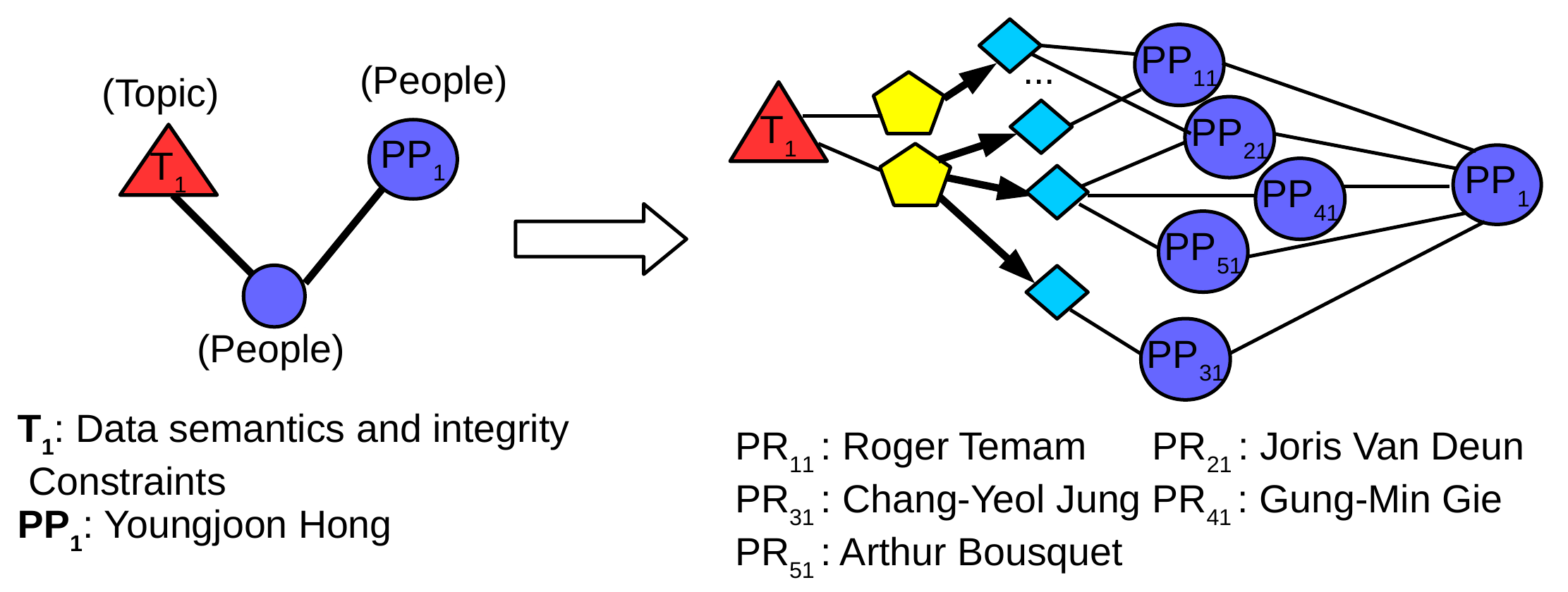}} 

	\subfloat[]{\includegraphics[width= 2.6in, height=1.2in]{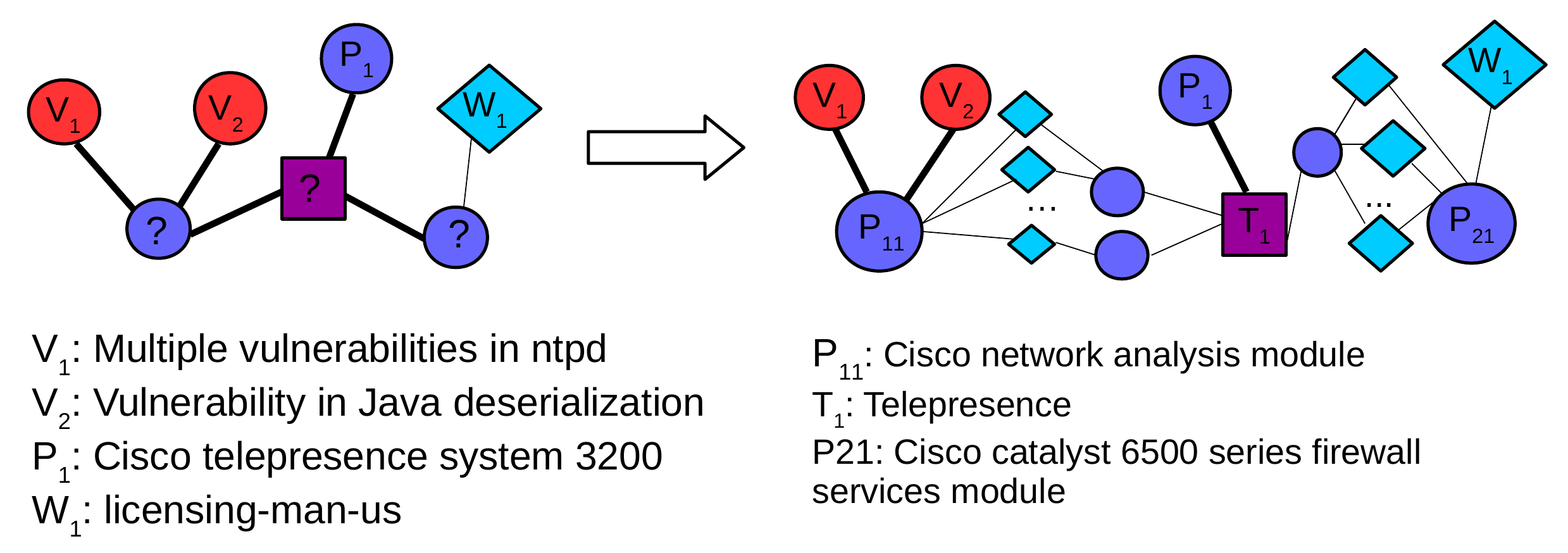}}
	\label{fig:nonStarQueryCiscoQuality} 
	\subfloat[]{\includegraphics[width= 2.6in, height=1.2in]{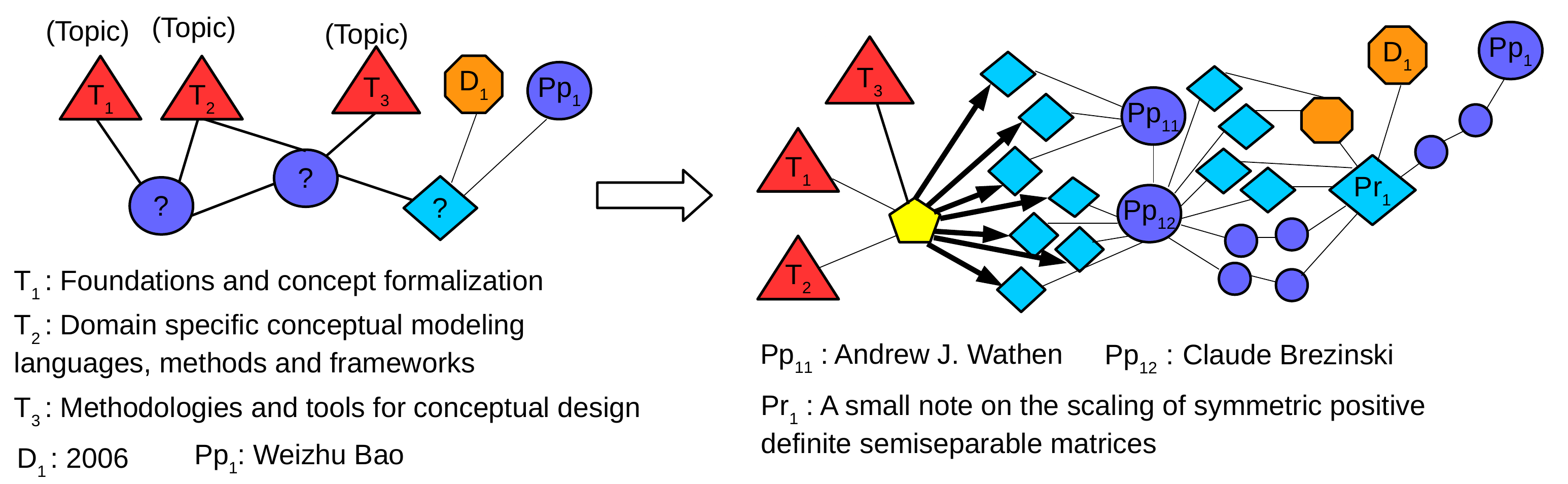}}
	\label{fig:nonStarQueryDblpQuality} 
		
	\caption{Graph queries and their results (a) Hierarchical star query graph in Cisco -- Query the products with Vulnerability $V_1$ and used Technology $T_1$, and its top-$5$ results in order. (b) Mixed star query graph in Dblp -- Query authors that work on Topic $T_1$ and cooperate with Person $Pp_1$, and its top-$5$ results in order. (c) Mixed general query graph in Cisco -- Query a product that has Vulnerability $V_1$ and $V_2$, another product that belongs to a workgroup $W_1$, and their common technology also used in a product $P_1$, and the top-$1$ result. (d) Mixed general query graph in Dblp -- Query an author that works on Topic $T_1$ and $T_2$ , and cooperates with another query person that works on Topic $T_3$, and published a paper coauthored with Person $Pp_1$ in year $Di_1$, and its top-$1$ result.} 
	\label{fig:qualityGraphExamples}
	\vspace*{-4.8mm}
\end{figure}

\vspace*{-1.2mm}
As mentioned earlier, a query graph can be classified as a hierarchical, mixed hierarchical or non-hierarchical query graph considering inheritance relations, and be a star query or general query graph based on query node numbers.
We show the results of hierarchical star query graphs and mixed general query graphs here. In each real dataset, one star query example and non-star query example results are shown in Fig \ref{fig:qualityGraphExamples}. Fig \ref{fig:qualityGraphExamples}a shows the hierarchical star query with all specific nodes as attaching nodes and the query node as an inherited node, and the top-$5$ query results are found in Cisco data. As seen in the results, different inherited versions of Cisco WebEx meeting server products are queried with higher matching scores. Fig \ref {fig:qualityGraphExamples}b displays different authors with publication papers in a journal and working on the same topic, which is verified to be reasonable online. As more complex non-star queries shown in Fig \ref{fig:qualityGraphExamples}c and \ref{fig:qualityGraphExamples}d with each top-$1$ result, our algorithm GQH can also provide the top relevant query answers.

\vspace*{-2.8mm}
\subsection{Comparisons of Query Quality}
\vspace*{-1.1mm}
Existing state-of-the-art algorithms for graph query generally do not consider the hierarchical inheritance relations for querying. It is not meaningful to compare the results of query algorithm with different problem formulations directly. However, our paper proposes the query with hierarchical inheritance relations extending the query algorithm from the Jin et al. \cite{jin2015querying} (GraB Query), which does not utilize hierarchical inheritance relations. NeMa \cite{khan2013nema} is a recent and classical method for neighborhood-based query also without considering hierarchical inheritance relations. Therefore, we compare the effectiveness of our query algorithm with the graph query based on GraB query and NeMa query algorithms.

\vspace*{-2.8mm}
\subsubsection{Comparison with GraB Query Algorithm}
First, we show the quality of GraB's query result on the database with hierarchical inheritance relations on Cisco data and Extended Dblp data.
We compare with our algorithm GQH based on the example of the query in Fig \ref{fig:qualityGraphExamples}a and show the result. 
Table \ref{compareJinResultInCisco} shows top-$5$ results of comparison with GraB's Query algorithm. GQH shows the possible ``Cisco WebEx meeting server version" inheritances as more potential candidates than GraB's query algorithm, with three different number of matched candidates. This is because GraB's Query algorithm only considers the node types and shortest distances as metrics for ranking.

\vspace*{-4.1mm}
\subsubsection{Comparison with NeMa Query Algorithm}
NeMa in \cite{khan2013nema} uses nodes' label and neighborhood similarity in small hops to find the top matched subgraphs.  We compare the query quality with our algorithm GHQ for the query in Fig \ref{fig:qualityGraphExamples}a, and shows the result in Table \ref{compareNeMaResultInCisco}. It shows top-$5$ results of comparison with GraB Query algorithm. NeMa uses matching cost which measures the cost of matched subgraph with the query graph. The smaller the cost, the better the matching. ``---" indicates no matching result found, only top-$3$ results are returned. It also does not return the hierarchical ``Ciso WebEx meeting server" answers. Because it limits the maximum hops of its visits and does not consider the hierarchical inheritance,  which leads to a smaller structural difference but fewer potential matches.

\vspace*{-1.8mm}
\begin{table}[!ht]
	\centering
	\caption{Comparison with GraB algorithm in Cisco}
	\label{compareJinResultInCisco}
	\begin{tabular}{p{0.6cm}p{4.2cm}p{1.3cm}p{4.2cm}p{1.3cm}}
		\cline{2-5}
		& \multicolumn{2}{l|}{Query result in GQH} & \multicolumn{2}{l}{Query result in GraB } \\ \hline
		Rank &  \hspace{0.1in} Node                                     & \hspace{0.1in} Score  &  \hspace{0.1in} Node       & Score  \\ \hline
		1 & Cisco WebEx meetings server versions 0.2    &  \hspace{0.1in}1.7906           & Cisco WebEx meetings server versions 1.x     &   \hspace{0.1in} 1.7186           \\ \hline
		2   & Cisco WebEx meetings server versions 1.x    & \hspace{0.1in} 1.7100 & Cisco WebEx meetings server versions 2.x &  \hspace{0.1in} 1.7186 \\  \hline
		3   & Cisco WebEx meetings server versions 2.x     & \hspace{0.1in} 1.7100 & Easy vpn & \hspace{0.1in} 1.7186 \\  \hline
		4   &Cisco 12000 series spa interface processors running Cisco ios software   &  \hspace{0.1in} 1.7015  & Cisco unified ip phone  &  \hspace{0.1in} 1.7015 \\  \hline
		5   &Cisco xr 12000 series engine 3 line cards     & \hspace{0.1in} 1.7015 & Catalyst 6000 supervisor module & \hspace{0.1in} 1.7015 \\  \hline
	\end{tabular}
	\centering
	\caption{Comparison with NeMa algorithm in Cisco}
	\label{compareNeMaResultInCisco}
	\begin{tabular}{p{0.5cm}p{4.3cm}p{1.2cm}p{4.3cm}p{1.2cm}}
		\cline{1-5}
		& \multicolumn{2}{l|}{Query result in GQH} & \multicolumn{2}{l}{Query result in NeMa} \\ \hline
		Rank & \hspace{0.1in} Node                                     & \hspace{0.1in} Score  & \hspace{0.in} Node            & Cost  \\ \hline
		1    & Cisco WebEx meetings server versions 0.2    & \hspace{0.3in} 1.7906    & Cisco WebEx meetings server version 0.2    &  \hspace{0.1in} 2.0   \\  \hline
		2   & Cisco WebEx meetings server versions 1.x   &  \hspace{0.1in}1.7100 & Cisco ASA Series show running-config prior to 7.2.1 & \hspace{0.1in}  2.95833 \\  \hline
		3   & Cisco WebEx meetings server versions 2.x     &  \hspace{0.1in}1.7100 & Cisco ASA Series show running-config between 7.2.1 and 8.4 &  \hspace{0.1in} 2.95833 \\  \hline
		4   &Cisco 12000 series spa interface processors running Cisco ios software     &  \hspace{0.1in} 1.7015  &  ---  & --- \\  \hline
		5   &Cisco xr 12000 series engine 3 line cards     &  \hspace{0.1in} 1.7015  & --- & ---  \\  \hline
	\end{tabular}
	\vspace*{-1.8mm}
\end{table}

\vspace*{-1.8mm}
\subsection{Efficiency of Graph Query}
\vspace*{-1.8mm}
Our GQH algorithm mainly focuses on improving the quality of query and we also use a different implementing platform and  programming language from GraB and NeMa algorithms, thus comparing the running time directly to them is not meaningful. Therefore, we test the efficiency of our GQH algorithm itself. We evaluate the efficiency with different top-$k$ values, query graph size and data graph size. For each different test, we keep the one testing parameter varied and the other unchanged. Each experiment is done 20 times and we get the average runtime with different parameters.

\begin{figure}[!ht]
	\subfloat[Synthetic]{\includegraphics[width= 0.90in, height=1.0in]{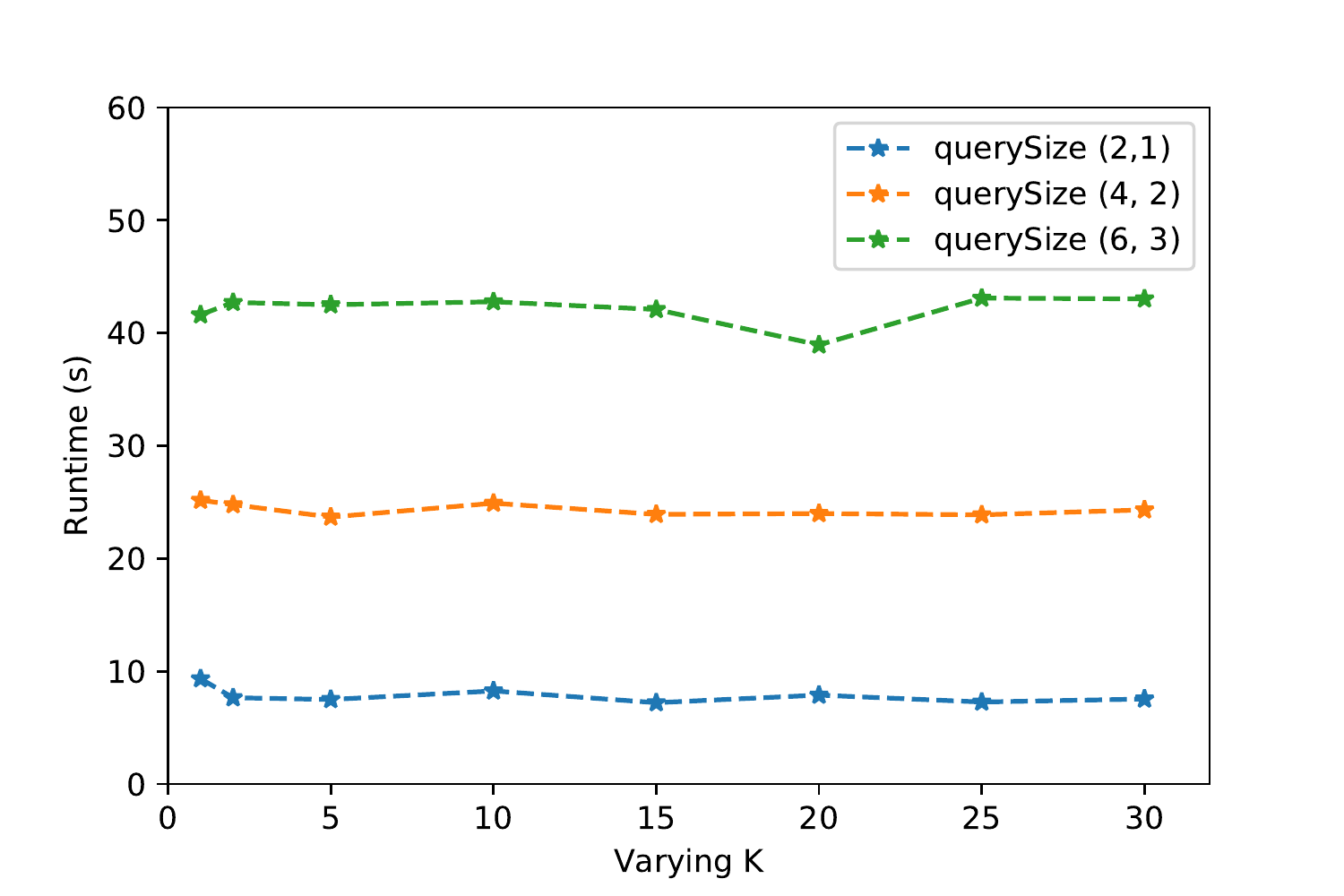}}
	\label{fig:varingTopKRuntimeSyntheticData}
	\subfloat[Cisco]{\includegraphics[width= 0.90in, height=1.0in]{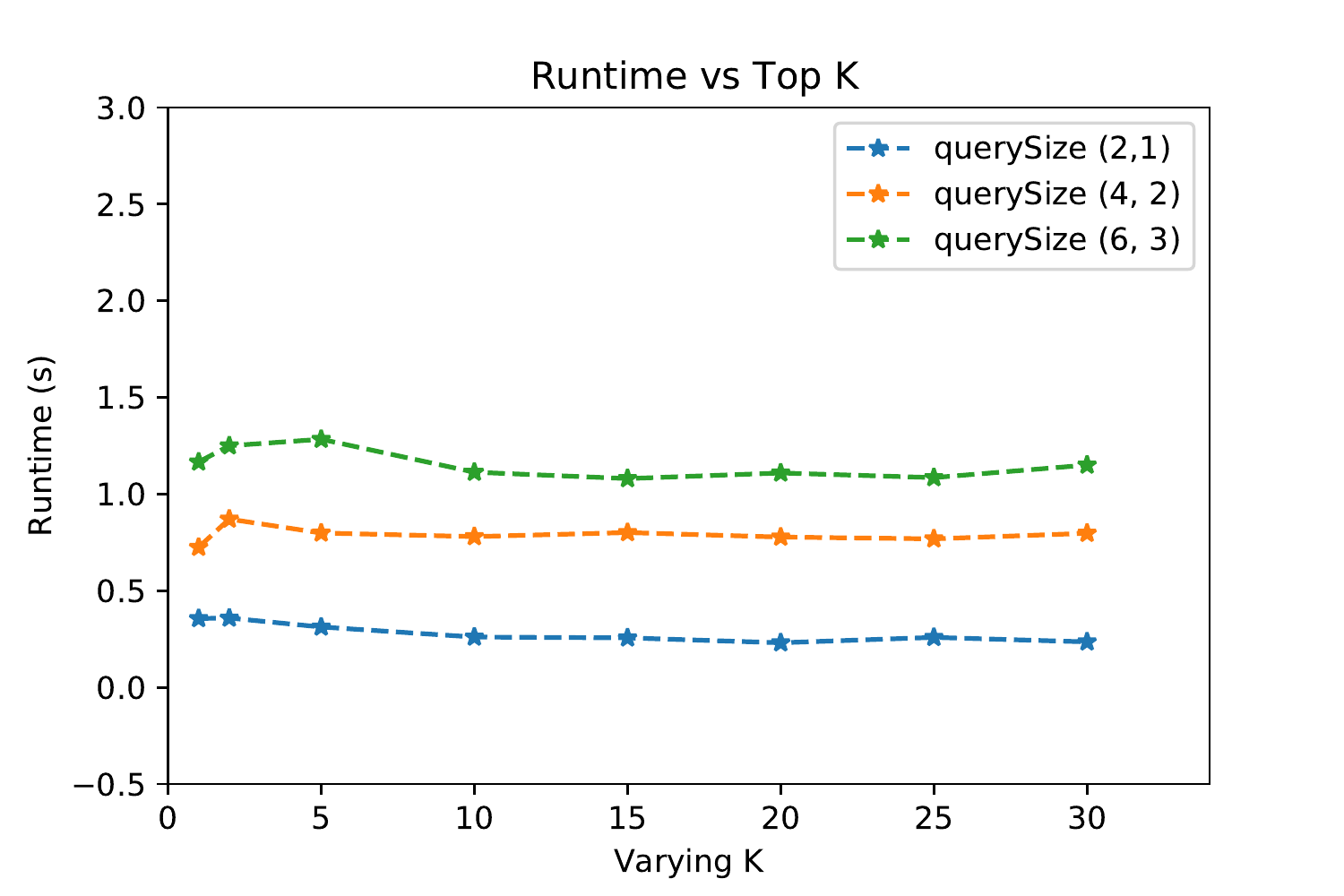}}
	\label{fig:varingTopKRuntimeSyntheticData}
	\subfloat[Dblp]{\includegraphics[width= 0.90in, height=1.0in]{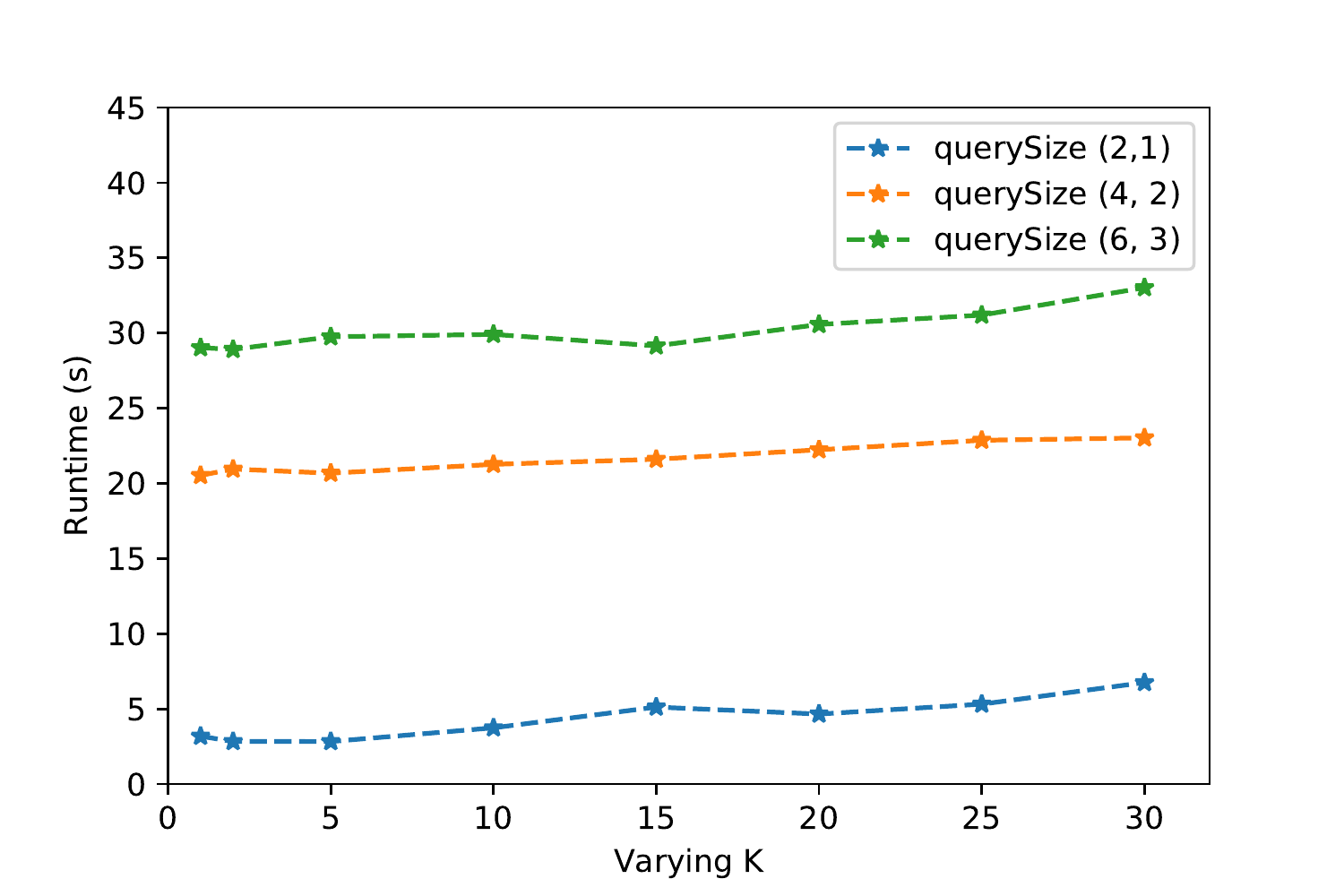}}
	\label{fig:varingTopKRuntimeDblpData}
	\subfloat[Synthetic]{\includegraphics[width= 0.98in, height=1.0in]{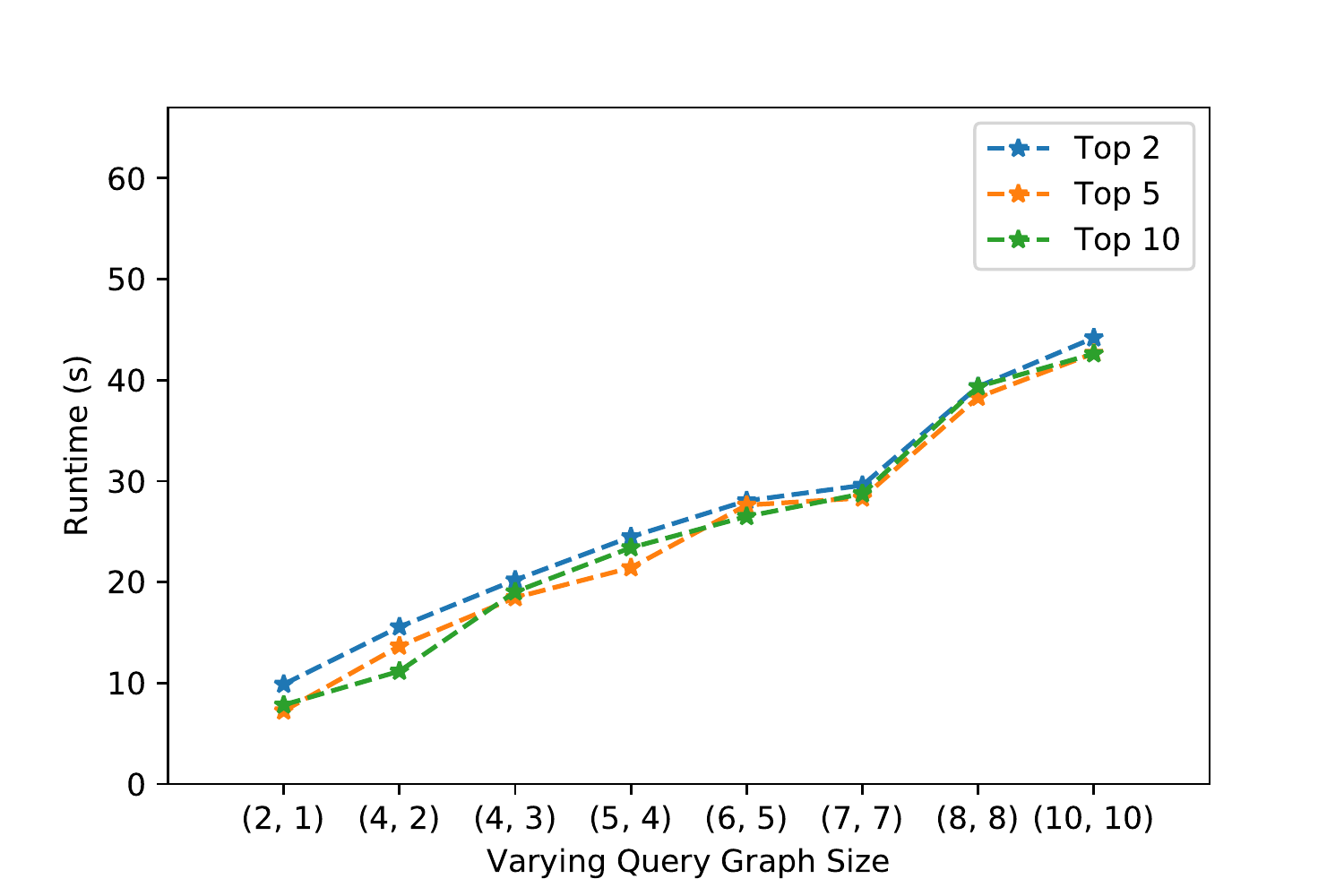}}
	\label{fig:varyingQuerySizeRuntimeSyntheticData}
	\subfloat[Cisco]{\includegraphics[width= 0.98in, height=1.0in]{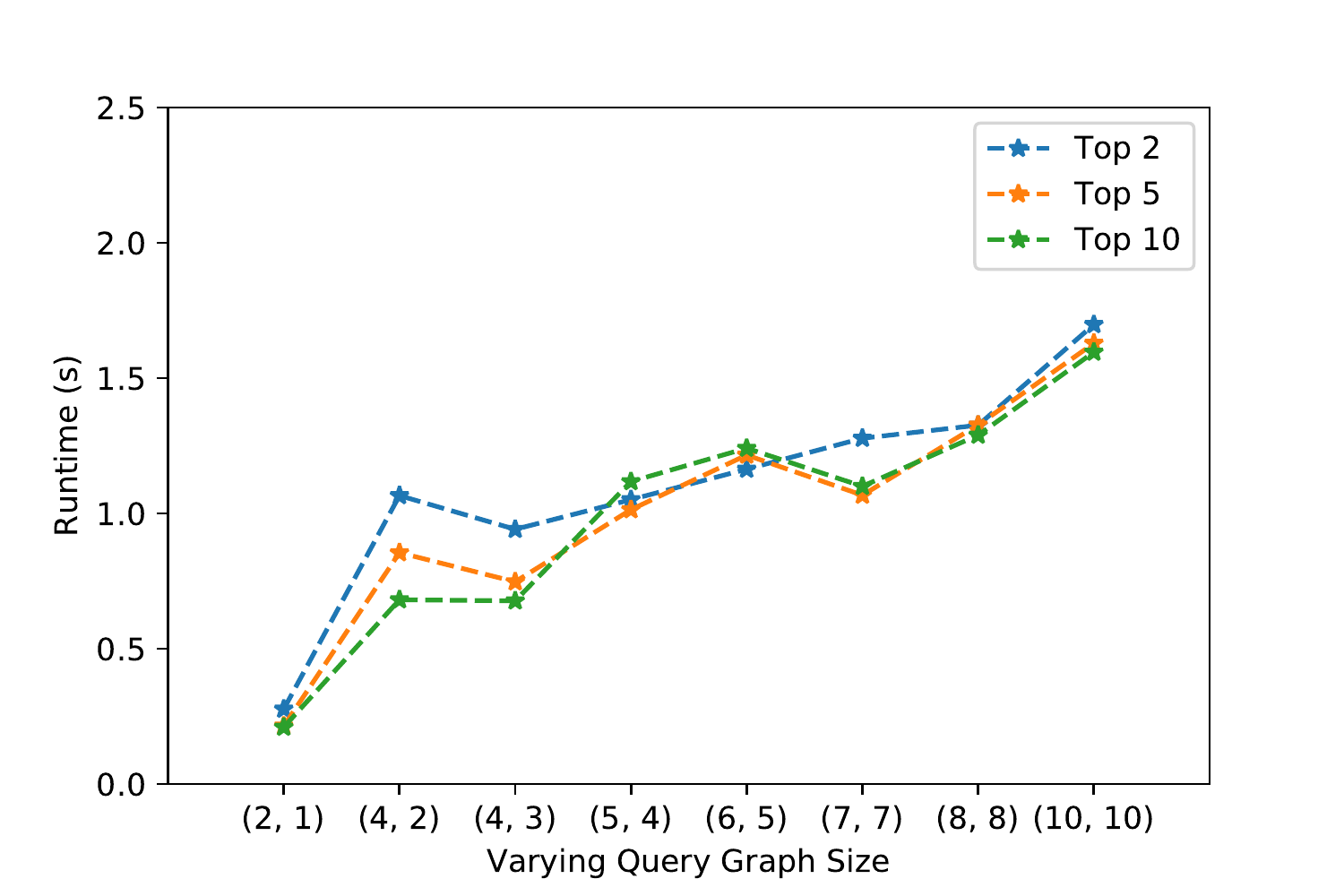}}
	\label{fig:varyingQuerySizeRuntimeCiscoData}
	
	\subfloat[Dblp]{\includegraphics[width= 1.0in, height=1.0in]{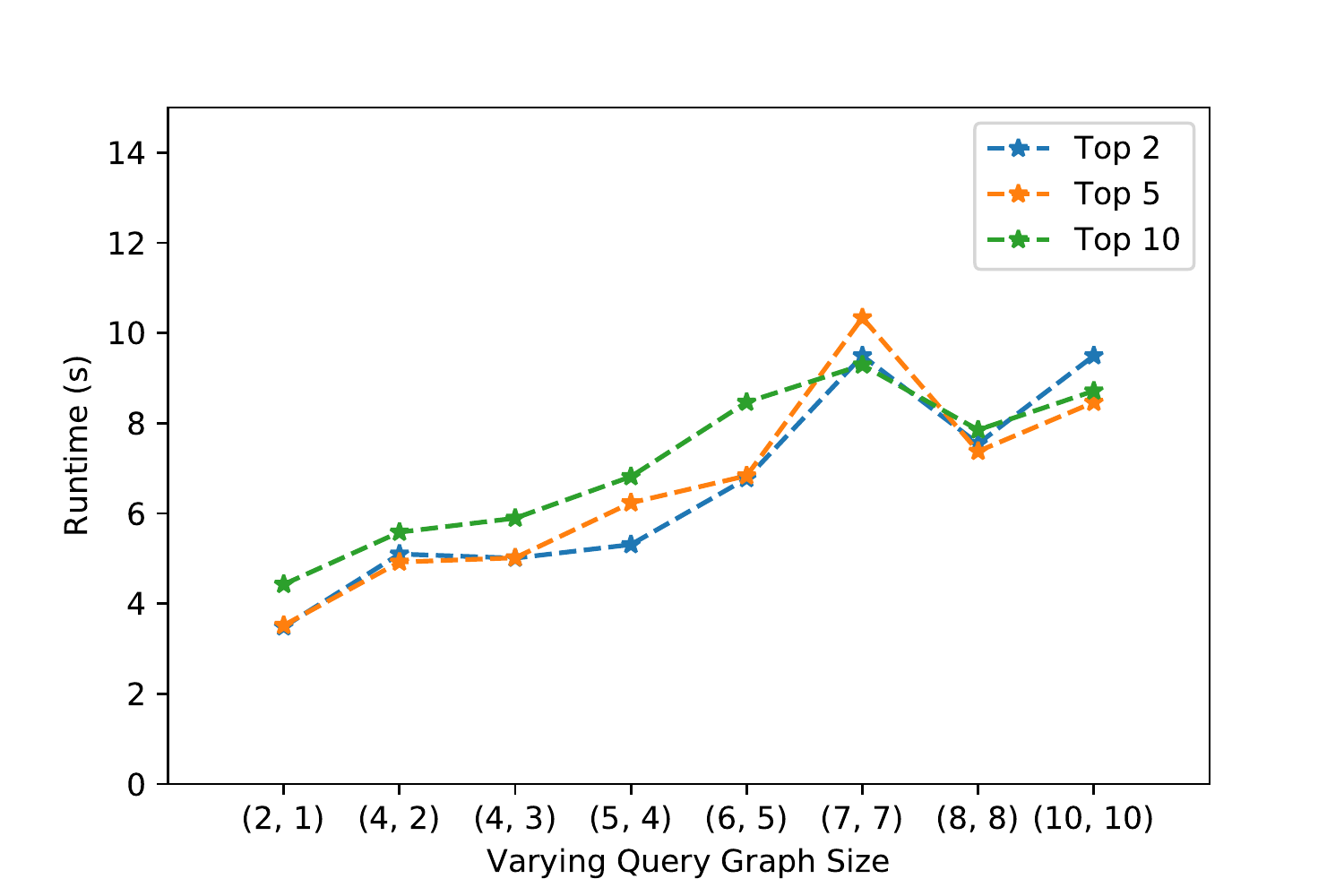}}
	\label{fig:varyingQuerySizeRuntimeDblpData}
	\subfloat[Synthetic]{\includegraphics[width= 0.91in, height=1.0in]{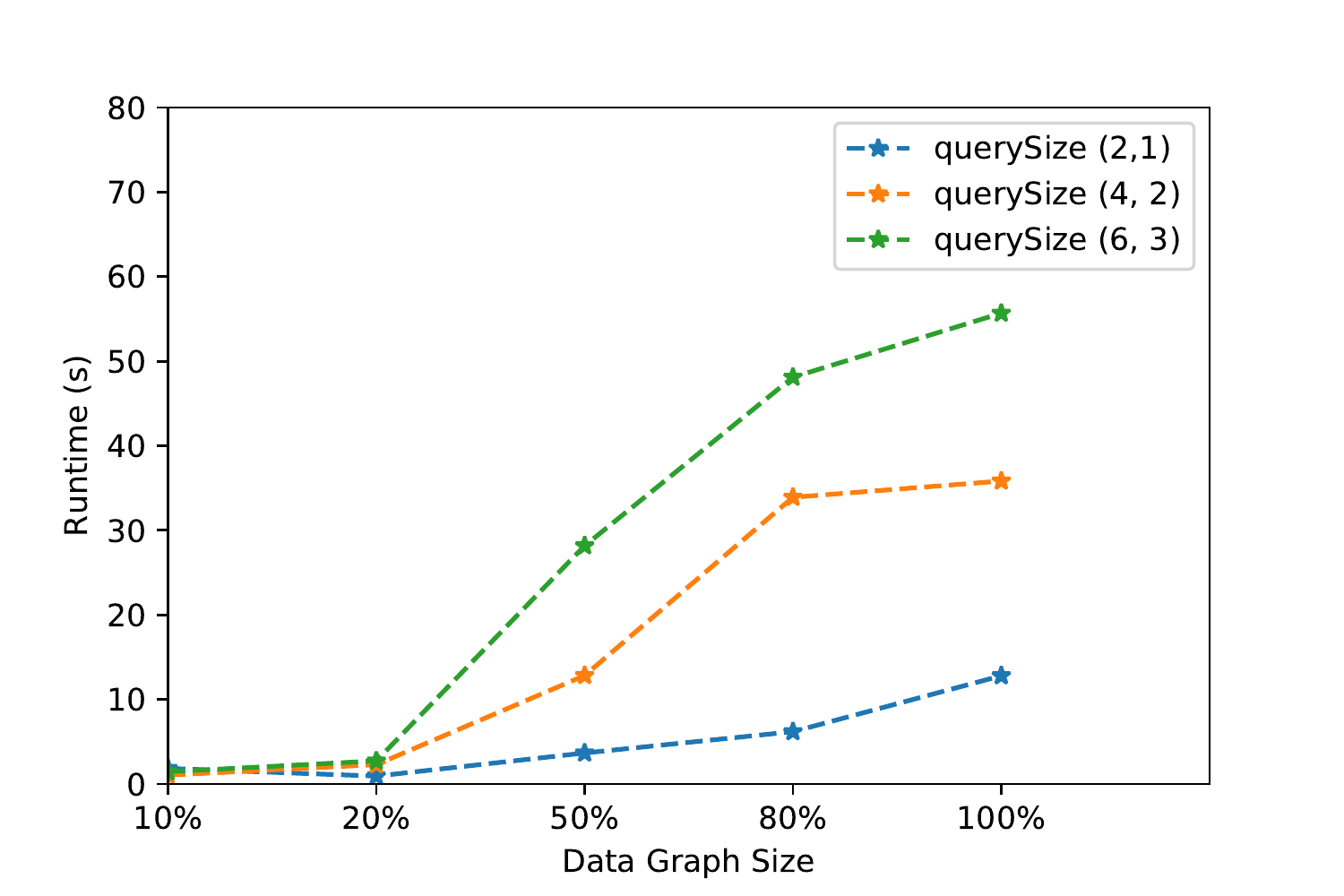}}
	\label{fig:varyingDataGraphSizeRuntimeSyntheticData}
	\subfloat[Cisco]{\includegraphics[width= 0.91in, height=1.0in]{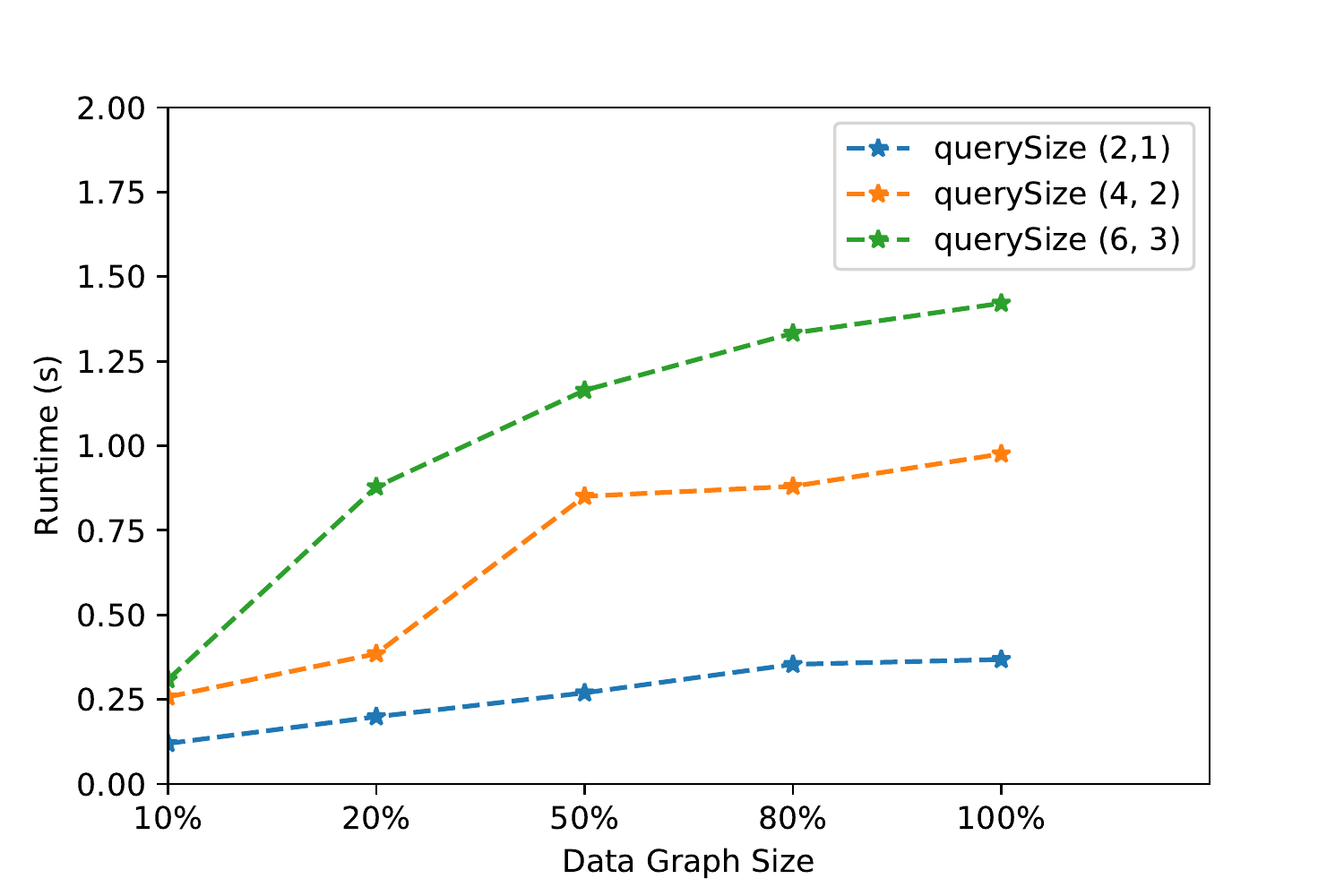}}
	\label{fig:varyingDataGraphSizeRuntimeCiscoData}
	\subfloat[Dblp]{\includegraphics[width= 0.91in, height=1.0in]{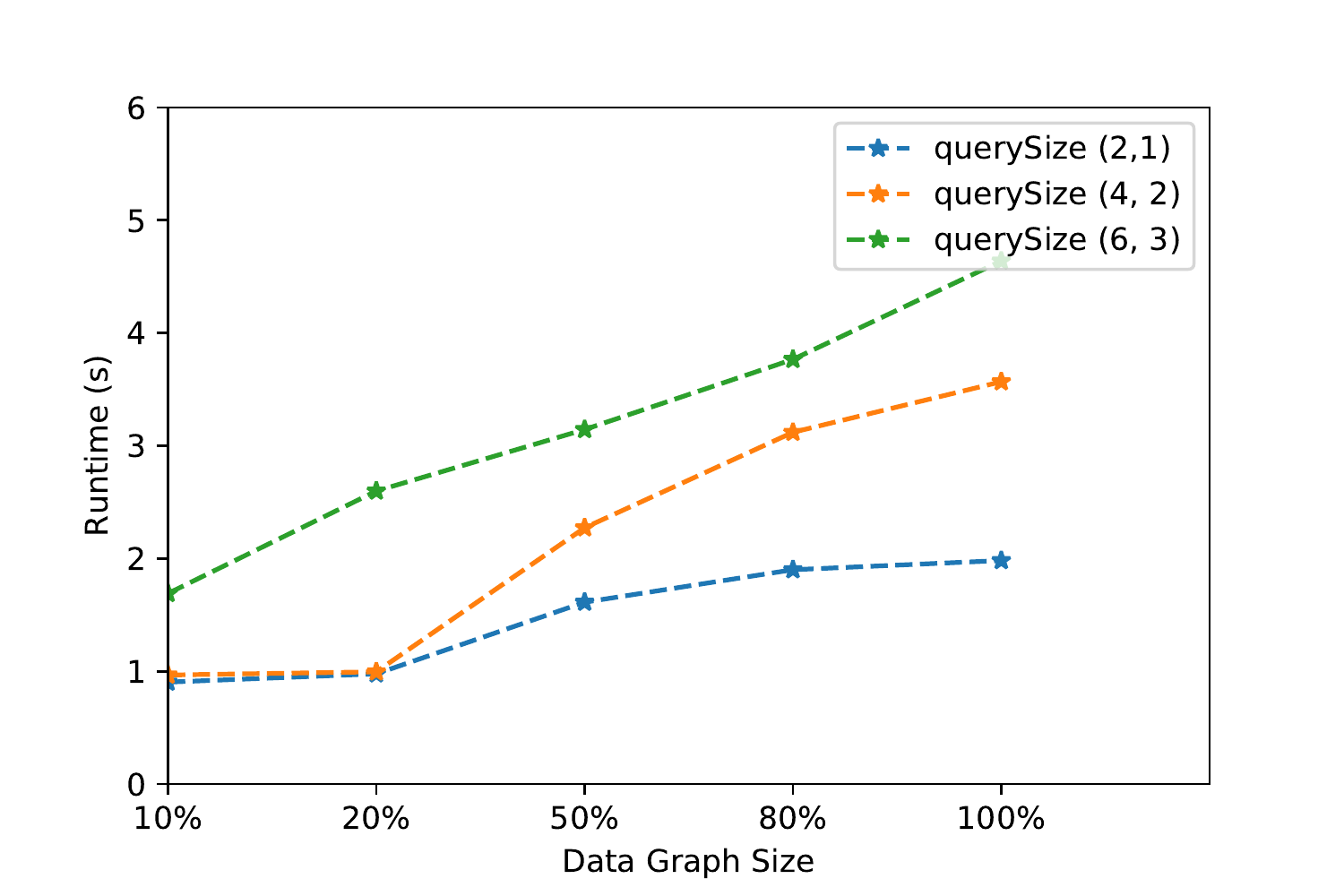}}
	\label{fig:varyingDataGraphSizeRuntimeDblpData}
	\subfloat[Synthetic]{\includegraphics[width= 0.94in, height=1.0in]{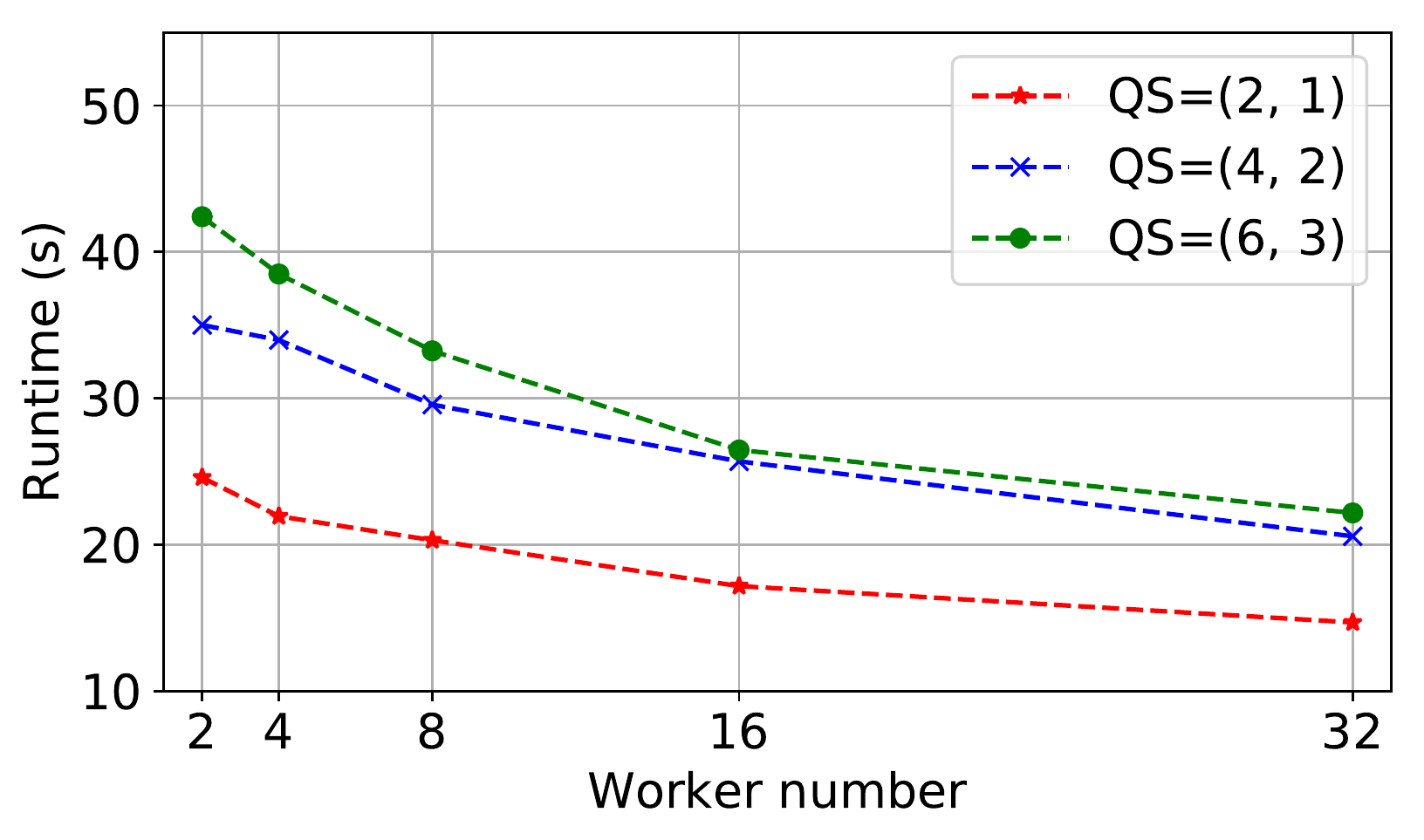}}
	\label{fig:varyingWorkerNumberSyntheticData}
	\caption{Efficiency and scalability}  
	\label{fig:efficencyScalableTest}
	\vspace*{0.8mm}
\end{figure}

\textbf{Varying $k$:}
To check how our algorithm scales with different querying $k$, we examine the average runtime for the different top-$k$ from 1, 2, 5 to 30 in Fig \ref{fig:efficencyScalableTest}a--c. Three different query sizes [2,1], [4,2], and [6,3] are fixed. It shows the runtime is basically sublinear no matter the $k$ value. This is because the complexity degrees of graphs lead to more than designated top-$k$ answered before the termination of iterations. We only fetch the top-$k$ candidates from all the obtained candidates.

\vspace*{-0.2mm}

\textbf{Varying Query Graph Size} To check how our algorithm scales with different query graph size, we examine the average runtime for the different query sizes. The query size is defined as a tuple(specific node number, query node number). We select (2,1), (4,2) to (10,10) shown  in Fig \ref{fig:efficencyScalableTest}d-f with top 2, 5, and 10 used. It shows the running time is basically sublinear with the increasing of query size.

\textbf{Varying Data Graph size:}
We test the query time with varying data graph size. We randomly and accumulatively extract subgraph from the original data graph for different nodes number, covering $10\%$, $20\%$,  $50\%$, $80\%$, and $100\%$. We measure 3 different query sizes in this scene to show the trend of query time for different graph data size. As shown in Fig \ref{fig:efficencyScalableTest}g-i, the query time also increases sublinearly with the increasing of graph data size.

\textbf{Scalability on Multiple Machines:}
To test the scalability of our algorithm on multiple machines, we test it on Google Cloud Platform to see the average runtime trends with increasing worker machines deployed. We use 3 different query size, with one master and increased worker machines from 2,4,...,32. Fig \ref{fig:efficencyScalableTest}j shows that the running average time decreases sublinearly with numbers of workers. 
\vspace*{-5.4mm}

\section{Related Work}
\vspace*{-1.2545mm}
\label{relatedWork}
There exist several classification categories for graph query. Based on user inputs, it can be classified as keywords query and structured query \cite{khan2013nema,roy2015fast}. Based on query answers, it includes exact match and inexact match \cite{khan2011neighborhood,lee2012depth,mongiovi2010sigma}. Based on matching techniques, it mainly contains indexing-based query and graph-traversal-based query for distance, neighbor and random walk \cite{fujiwara2012fast,zheng2013efficient}. Our algorithm focuses on the top-$k$ inexact match for structured graph query with hierarchical inheritance relations. (1) \textbf{Structured Graph Query:}
Various techniques have been proposed for structured graph query. Recent works allow users to express their own input query as a structured query graph and do the graph traversal based on node and path similarity for matching. For example, NeMa \cite{khan2013nema} and SLQ \cite{yang2014schemaless} consider different similarity transformations for node to match query graph and subgraphs in a data graph. Su et al. \cite{su2015exploiting} consider the graph query based on user relevances to further improve the query quality. Some works \cite{roy2015fast,du2017first} consider the multiple attributes of nodes for graph query. Jin et al. \cite{jin2015querying} propose a specified ranking function for structured graph query with specific nodes to find answer nodes. Most of them use indexing which takes large spaces, or graph traversal with only two dimensions of node and edge similarities. We consider one more dimension of hierarchical inheritance relations for effective queries. (2) \textbf{Top-$k$ Graph Query:} Top-$k$ graph query tries to get top-$k$ matched answers for the graph query. The common practice for top-$k$ search is to use threshold algorithms to find the top matches by traversing sorted node/edge list \cite{fagin2003optimal}. They require precomputed and sorted lists to derive the bounds. Recent top-$k$ query have been studied in \cite{cheng2013top,gupta2014top}. Yang et al. \cite{yang2016fast} consider the STAR-query structure and top-$k$ ranked join for general graph query, but the matches are limited to answer subgraphs with paths of bounded length. Our algorithm considers top-$k$ general graph with an efficient ranking and bounded-based solution without the limitation of path lengths for hierarchical relation inheritance.

\vspace*{-1.8mm}
\section{Conclusion}
\vspace*{-1.8mm}
\label{conclusion}
We consider an additional dimension of hierarchical inheritance relations on real-world  heterogeneous information networks for graph query. The problem is reformulated with hierarchical inheritance relations, and we propose a graph query algorithm based on that for star-query and general graph query. With the bounding-based techniques, our algorithm can effectively capture hierarchical inheritance relations on information networks for better query answers and competitive performances are also achieved.

\vspace*{-2.8mm}
%
%
 \bibliographystyle{splncs04}
 
 \bibliography{reference}

\par\noindent 
\parbox[t]{\linewidth}{
\noindent\parpic{\includegraphics[height=1.5in,width=1in,clip,keepaspectratio]{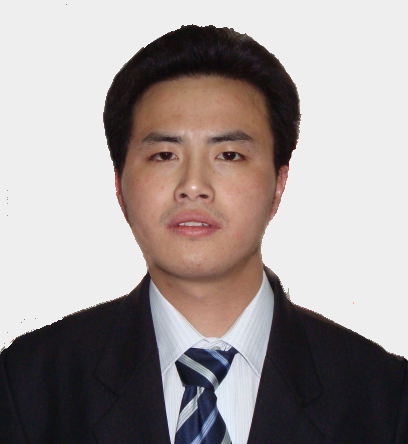}}
\noindent {\bf Fubao Wu}\
 received his Bachelor degree in Electronic Information Engineering from Northeastern 
University, China in 2008 and Master of Engineering in Electronic Science and Technology from University of Science and Technology of China in 2011. He is currently pursuing his Ph.D. degree
in Electrical and Computer Engineering at University of Massachusetts, Amherst. His research interests are data analytics, graph analytics and video analytics.}
\vspace{4\baselineskip}

\par\noindent 
\parbox[t]{\linewidth}{
\noindent\parpic{\includegraphics[height=1.5in,width=1in,clip,keepaspectratio]{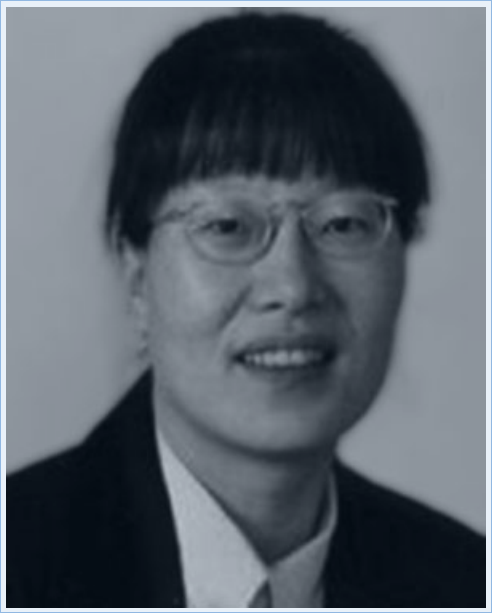}}
\noindent {\bf Lixin Gao}\
 received the Ph.D. degree in computer science from the University of Massachusetts at Amherst, in 1996. Now she is a professor of electrical and computer engineering with the University of Massachusetts at Amherst. Her research interests include social networks, Internet routing, network virtualization and cloud computing. Between May 1999 and January 2000, she was a visiting researcher in AT\&T Research Labs and DIMACS. She was an Alfred P. Sloan fellow between 2003-2005 and received an NSF CAREER Award in 1999. She won the best paper award from IEEE INFOCOM 2010 and ACM SoCC 2011, and the test-of-time award in ACM SIGMETRICS 2010. She received the Chancellors Award for Outstanding Accomplishment in Research and Creative Activity in 2010. She is a fellow of the IEEE and ACM.}
\vspace{4\baselineskip}

\end{document}